\pdfoutput=1
\documentclass[preprint,journal]{vgtc}

\graphicspath{{figures/}{pictures/}{images/}{./}}

\usepackage{mathptmx}

\setitemize{noitemsep,topsep=0pt,parsep=0pt,partopsep=0pt}

\onlineid{1438}

\vgtccategory{Research}
\preprinttext{To appear in IEEE Transactions on Visualization and Computer Graphics.}
\renewcommand{\manuscriptnotetxt}{\vspace{.25in}}

\title{\textsc{CrossAtlas}: Evaluating Projection Techniques for Spatial Referencing in Cross-Reality Collaboration}

\author{%
  \authororcid{Haoyang Yang}{0000-0002-0566-0169},
  \authororcid{Chenyang Zhang}{0009-0003-1116-4895},
  \authororcid{Elliott H. Faa}{0009-0002-8698-0961},
  \authororcid{Weijian Liu}{0009-0004-4771-2539},
  \authororcid{Lily Seika Chisholm}{0009-0006-7120-3747},\texorpdfstring{\\}{ }
  \authororcid{Benjamin Lee}{0000-0002-1171-4741},
  \authororcid{David Saffo}{0000-0001-9515-048X},
  \authororcid{Feiyu Lu}{0000-0002-1939-9352},
  \authororcid{Blair MacIntyre}{0000-0002-5357-2366},
  and 
  \authororcid{Yalong Yang}{0000-0001-9414-9911}
}

\authorfooter{
    \item
  	Haoyang Yang, Chenyang Zhang, Elliott H. Faa, Weijian Liu, Lily Seika Chisholm, and Yalong Yang are with Georgia Tech. 
    E-mail: \{alexanderyang, chenyang.zhang, efaa3, wliu430, lchisholm30, yalong.yang\}@gatech.edu
    \item
  	Benjamin Lee is with the University of Stuttgart.
  	E-mail: benjamin.lee@visus.uni-stuttgart.de
    \item
  	David Saffo is with JPMorgan Chase.
  	E-mail: david.saffo@jpmchase.com
    \item
  	Feiyu Lu is with Virginia Tech.
  	E-mail: feiyulu@vt.edu
    \item
  	Blair MacIntyre is with Northeastern University.
  	E-mail: b.macintyre@northeastern.edu
}

\teaser{
  \centering
  \vspace{0.5em}
  \includegraphics[width=\linewidth]{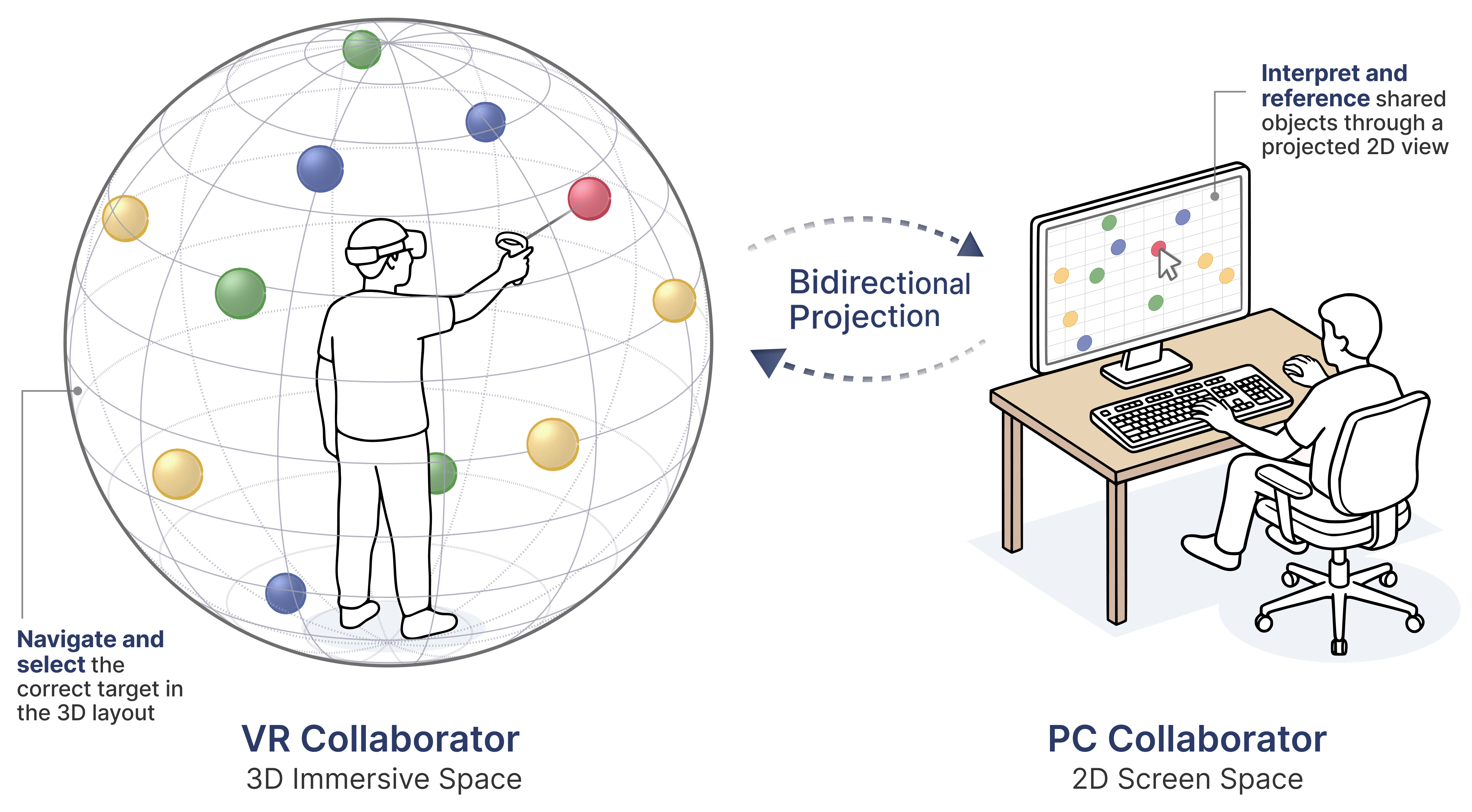}
  \vspace{-1.75em}
  \caption[\textsc{CrossAtlas} cross-reality collaboration platform]{
  \textsc{CrossAtlas} supports PC--VR collaboration for spatial referencing tasks. The VR collaborator (left) interacts with a 3D object layout in VR, whereas the PC collaborator (right) views its projected 2D representation. Together, they use spatial references to identify target objects across views, with bidirectional projection maintaining correspondence between the two representations.}
  \label{fig:teaser}
  
}

\abstract{
Cross-reality collaboration increasingly connects immersive and desktop users within synchronized workspaces, yet little is known about how bidirectional projection techniques between immersive 3D layouts and desktop 2D views influence communication. 
Spatial referencing depends on shared spatial understanding, but different mappings preserve and distort geometric relationships in different ways, altering perceived adjacency, orientation, and coverage across collaborators' views. 
We present \textsc{CrossAtlas}, a synchronized PC--VR collaboration platform that integrates multiple bidirectional projection techniques, including three planar projection variants and \textit{equirectangular}, a spherical projection variant, across layouts of varying curvature.
In a controlled study with 24 dyads, collaborators completed spatial referencing tasks under different projection--layout conditions while we collected performance and subjective measures. 
Our results show that projection choice strongly shaped collaboration, with the spherical variant often outperforming planar projections and remaining robust across object layouts. 
}
 

\usepackage[svgnames]{xcolor}
\usepackage{xspace}

\definecolor{agreen}{RGB}{74, 198, 148}
\definecolor{purple}{RGB}{158, 62, 177}
\definecolor{darkpurple}{RGB}{170, 70, 210}
\definecolor{aqua}{RGB}{87, 180, 181}
\definecolor{lightblue}{RGB}{72, 123, 232}
\definecolor{hotpink}{RGB}{255, 83, 115}
\definecolor{teal}{RGB}{90, 200, 250}
\definecolor{linkColor}{RGB}{0, 128, 229}
\definecolor{lightgreen}{RGB}{33, 222, 128}
\definecolor{almostBlack}{RGB}{60,60,60}

\definecolor{red}{RGB}{255, 0, 0}
\definecolor{green}{RGB}{0, 128, 0}
\definecolor{yellow}{RGB}{255, 192, 0}
\definecolor{cyan}{RGB}{0, 255, 255}
\definecolor{lightgray}{gray}{0.95}
\definecolor{grayborder}{gray}{0.5}
\definecolor{gray}{gray}{0.75}
\definecolor{orange}{RGB}{236, 107, 44}
\definecolor{lightorange}{RGB}{255, 223, 186}
\definecolor{lightpurple}{RGB}{202,58,126}
\definecolor{benign_purple}{RGB}{150,150,150}
\definecolor{adv_orange}{RGB}{212,64,57}

\definecolor{pca}{HTML}{E69F00}
\definecolor{horizontal}{HTML}{009E73}
\definecolor{vertical}{HTML}{D55E00}
\definecolor{equirectangular}{HTML}{56B4E9}

\usepackage{wrapfig}
\usepackage{gensymb}

\definecolor{niceblue}{RGB}{56, 116, 203}

\AtBeginDocument{%

  \crefname{figure}{fig.}{figs.}
  \Crefname{figure}{Fig.}{Figs.}
  \crefname{equation}{eq.}{eqs.}
  \Crefname{equation}{Eq.}{Eqs.}
  \crefname{section}{\S}{\S}
  \Crefname{section}{\S}{\S}
  \crefname{subsection}{\S}{\S}
  \Crefname{subsection}{\S}{\S}
  \crefname{subsubsection}{\S}{\S}
  \Crefname{subsubsection}{\S}{\S}
  \creflabelformat{equation}{#2\textup{#1}#3}
  \hypersetup{
    linkcolor=niceblue,
    citecolor=niceblue,
    urlcolor=niceblue,
    filecolor=niceblue
  }
}

\newcommand{\figpart}[1]{\textcolor{niceblue}{#1}}

\begin{document}

\firstsection{Introduction}

\maketitle

From virtual and augmented reality headsets to high-resolution desktop displays, the rapidly evolving landscape of display and interaction technologies offers tremendous opportunities for creating innovative human-computer interaction experiences. 
Immersive environments, in particular, can transform the space surrounding a user into an expansive workspace in which information, applications, and documents extend beyond the boundaries of a single display.
Prior work has explored this affordance through immersive ``Space to Think'' systems~\cite{lisle_evaluating_2020,lisle_sensemaking_2021}, VR window and application management~\cite{marguet_windowspace_2025}, immersive computational notebooks and dashboards~\cite{in_evaluating_2024,in2025exploring,batch_there_2020,lee_shared_2021}, collaborative document and layout organization~\cite{luo_where_2022}, and spatial co-design and planning systems~\cite{borowski_spatialstrates_2025}.
Across these applications, distributing information throughout 3D space supports sensemaking, spatial memory, and flexible exploration by enabling users to externalize and organize information within the surrounding environment~\cite{lisle_evaluating_2020,yang_litforager_2025,seraji_analyzing_2024,yang2020embodied}. 
Importantly, these benefits extend beyond single-user interaction.
In collaborative immersive analytics, embodied interaction and expansive spatial workspaces have been shown to facilitate collective sensemaking through natural manipulation, spatial externalization, and concurrent access to large information spaces~\cite{lee_shared_2021,cordeil_immersive_2017}.

However, not all collaborators may be able or willing to work in an immersive environment.
Differences in hardware availability, task role, physical ability, and user preference naturally give rise to scenarios in which team members work across heterogeneous platforms~\cite{brehmer2026challenges}. 
Such cross-platform collaboration offers unique advantages, as devices can mutually scaffold each other's weaknesses~\cite{horak_when_2018}: immersive environments provide embodied spatial exploration and expansive workspaces, while desktop systems contribute precision input, structured manipulation, and broad accessibility.
This paradigm is formalized as \textit{cross-reality collaboration}, in which collaborators situated at different positions along the reality--virtuality continuum~\cite{milgram_taxonomy_1994} share a synchronized interactive workspace~\cite{ens_revisiting_2019}.
A growing number of systems have developed techniques for synchronizing shared views and representations, supporting cross-platform awareness, and enabling workflow continuity between immersive and non-immersive interfaces~\cite{tong_towards_2023, saffo_through_2023}.

Despite this progress, a fundamental challenge remains underexplored. Existing cross-reality systems have primarily focused on interaction techniques and awareness mechanisms across devices~\cite{borowski_dashspace_2025, srinivasan_heedvision_2025}.
Far less attention has been paid to how the same spatial organization may be represented differently when adapted for immersive and desktop platforms, and how these differences shape collaboration.
For example, during collaborative exploration, an immersive user may rely on a surrounding 3D layout for spatial encoding, while a desktop collaborator views a screen-based representation designed for navigation, grouping, or annotation. 
Although both users engage with the same underlying data, it is presented through different spatial configurations.
As a result, collaborators may not interpret spatial relationships in the same way, making it harder to establish a shared frame of reference across views.

Effective cross-reality collaboration therefore depends heavily on spatial referencing. 
When one collaborator says ``the item next to the red cluster'' or ``above the landmark,'' the success of that reference depends on both partners interpreting the layout consistently.
If the 3D-to-2D transformation distorts adjacency, orientation, or relative position, references that are clear in one view may become ambiguous in the other, increasing miscommunication and degrading efficiency~\cite{enriquez_evaluating_2024}. 
Productive cross-reality collaboration thus requires perspective sharing---mutual awareness of how each collaborator perceives the shared space.
A key aspect of perspective sharing is \textit{spatial congruency}: the degree to which spatial relationships in one view map predictably onto the other~\cite{borowski_spatialstrates_2025,saffo_through_2023}.

The challenge of achieving \textit{spatial congruency} is fundamentally a bidirectional mapping problem. 
Because immersive and desktop users operate on the same workspace through different representations, the mapping must support both an interpretable 2D view and reliable correspondence back to the original 3D layout. 
In current practice, most cross-reality systems address this problem using \textbf{planar projections}, which map 3D content onto a single 2D plane~\cite{borowski_spatialstrates_2025, zhao_spatialtouch_2025, lee_design_2022}. 
Despite their conceptual simplicity and ease of implementation, planar projections often sacrifice surrounding spatial context by collapsing one spatial dimension and limiting directional coverage to a bounded field of view.
This limitation motivates the exploration of a broader projection design space for cross-reality systems. 
In particular, map-inspired \textbf{spherical projections} used in cartographic practice~\cite{lu_design_2025,yang2018maps} offer a compelling yet largely unexplored alternative for cross-reality space management.
Rather than depicting the scene from a single direction, spherical projections map positions around the user into a continuous 360\degree{} view, preserving angular coverage of the surrounding space. 
Although well established in cartography and panoramic media, they have not, to our knowledge, been evaluated against planar projections for cross-reality spatial referencing.
 
To address this gap, we developed \textsc{CrossAtlas}, a synchronized PC--VR platform for systematically evaluating how 3D-to-2D projection techniques influence collaboration across VR layout configurations. The name \textsc{CrossAtlas} reflects its role as an ``atlas'' of bidirectional mappings between shared 3D layouts and their 2D representations on the PC.
The system supports four projection techniques: three planar variants (\textit{horizontal}, \textit{vertical}, and \textit{PCA}) and one spherical variant (\textit{equirectangular}).
Using \textsc{CrossAtlas}, we conducted a controlled study with 24 dyads performing spatial referencing tasks of increasing complexity across 12 projection--layout conditions (eight per dyad) and collected performance metrics and subjective ratings. 
Our results suggest that projection choice influenced collaborative performance: the spherical variant performed better than \textit{horizontal} and \textit{PCA}, and it remained relatively robust across layout conditions, whereas each planar projection showed limitations under certain layouts.
By comparing planar projections with a map-inspired spherical alternative in collaborative tasks, our findings show that projection is not merely a visualization detail but a core communicative design factor in cross-reality collaboration.

In summary, we present the following major contributions:
\begin{itemize}[topsep=1pt, itemsep=0mm, parsep=3pt, leftmargin=9pt]

    \item \textbf{\textsc{CrossAtlas}, an open-source\footnote{\textsc{CrossAtlas}'s source code is publicly available at \url{https://github.com/AlexanderHYang/cross-atlas}.} cross-reality collaboration platform that integrates multiple bidirectional projection techniques}, including a map-inspired \textit{equirectangular} projection, within a shared framework for PC--VR spatial referencing.

    \item \textbf{A controlled empirical evaluation with 24 dyads} showing that projection choice significantly shapes collaborative spatial referencing performance, with the \textit{equirectangular} projection emerging as the most robust approach across layouts compared with planar projections.

\end{itemize}
\section{Related Work}

\smallskip
\noindent\textbf{Cross-Reality Collaboration.}
Cross-reality collaboration brings together collaborators situated at different positions along the reality--virtuality continuum~\cite{milgram_taxonomy_1994}, from conventional 2D desktop workstations to immersive AR/VR interfaces, into a shared interactive workspace~\cite{ens_revisiting_2019}. 
Research on cross-device and mixed-presence groupware has long identified coordination and shared awareness as central challenges in distributed work~\cite{gutwin_group_2004, badam_polychrome_2014, kim_hugin_2010, gou2026evaluating}. 
Cross-reality systems inherit these challenges while introducing additional complexity due to heterogeneous embodiment and view representations.

A growing body of work has developed systems that connect immersive and non-immersive platforms for collaborative tasks. 
Tong et al.\ investigated asymmetric collaborative visualization between VR and PC users, finding that the asymmetric PC--VR configuration could combine the strengths of both platforms without significant performance loss~\cite{tong_towards_2023}. 
Collaborative immersive analytics systems have further demonstrated that embodied interaction and expansive spatial work areas can benefit collective sensemaking through natural manipulation, spatial externalization, and simultaneous access to large information spaces~\cite{cordeil_immersive_2017, lee_shared_2021}. 
Comparative evaluations suggest that immersive settings can alter group dynamics and interaction patterns even when task accuracy is comparable to desktop baselines~\cite{billinghurst_collaborative_2002, butscher_clusters_2018}.

Despite this progress, most cross-reality systems prioritize awareness mechanisms---such as indicating gaze and position or providing overview representations---rather than preserving geometric equivalence across views~\cite{schroder_collaborating_2023, saffo_through_2023, srinivasan_heedvision_2025}. 
Desktop collaborators operate through fixed-frame 2D projections with indirect input, while immersive users act within egocentric 3D spaces where movement and orientation are tightly coupled to perception. This structural asymmetry means that even when objects share the same underlying coordinates, collaborators may perceive spatial relationships differently depending on how the 3D environment is projected into 2D.
Our work therefore presents a new focus: rather than awareness cues or interaction
techniques, we examine the projection transformation itself and how it shapes
collaborative spatial understanding.

\smallskip
\noindent\textbf{Spatial Awareness, Referencing, and Congruency.}
In immersive environments, users' spatial understanding is shaped by the structure of the layout around them.
Research on display curvature has shown that flat, semicircular, and circular arrangements produce different effects on spatial memory, navigation efficiency, and mental workload~\cite{liu_design_2020,liu_effects_2022}.
Liu et al.\ found that flat layouts supported more accurate spatial recall than full-circle layouts, while semicircular layouts offered a preferred compromise. 
Environmental landmarks and layout regularity further affect how users anchor and recall spatial positions~\cite{liu_investigating_2024}. 
These findings highlight how the geometry of the immersive environment shapes the spatial understanding that collaborators bring to cross-reality tasks.

In collaborative settings, partners often establish spatial references to direct each other's attention to locations in the shared workspace, using verbal cues such as ``next to the red chart'' or ``above the landmark''~\cite{heer_design_2008,clark_grounding_1991}. 
In homogeneous environments, these references are typically resolved within a shared coordinate system. 
In cross-reality settings, however, a reference grounded in immersive 3D space may map ambiguously onto a 2D view, requiring additional non-verbal cues, such as gaze or gesture tracking~\cite{wong_spatial_2025}, and increasing the risk of miscommunication.

At the group level, cross-reality collaboration requires collaborators to develop a shared understanding of the workspace. Saffo et al.'s ``eyes-and-shoes'' framework~\cite{saffo_through_2023} identifies four escalating levels of group awareness for cross-platform collaboration, with the highest, Level~4 (Perspective Sharing), enabling collaborators to understand how each partner perceives the space. 
A key factor in enabling perspective sharing is \textit{spatial congruency}: the degree to which two representations preserve perceived spatial relationships across views~\cite{borowski_spatialstrates_2025, wang_mrtransformer_2024,huang_surfshare_2024}.
Congruency is not purely geometric but perceptual and cognitive---mathematically valid 3D-to-2D mappings may differ substantially in how intuitively users can translate spatial relations.
Despite its importance, \textit{spatial congruency} has not been systematically studied as a function of the projection technique that bridges 3D and 2D views in cross-reality collaboration.

\smallskip
\noindent\textbf{Projection Techniques in Immersive Environments.}
Projection determines how immersive spatial content is externalized onto a 2D display.
In prior systems, planar projections are widely used for desktop and cross-reality representations because they show the scene from a single viewpoint within a bounded field of view~\cite{borowski_spatialstrates_2025, lee_design_2022, yang_origin-destination_2019, zhao_spatialtouch_2025}.
Planar projections map 3D content onto a flat surface, preserving local structure within a bounded view while sacrificing global continuity; the plane orientation determines which spatial axes are preserved or compressed. 
By contrast, spherical projections, widely used in cartography, map spherical coordinates onto a flat surface, preserving full 360\degree{} angular coverage at the cost of systematic distortion~\cite{ma_international_1998, snyder_map_1987}.
In single-user contexts, projection has primarily been studied as a rendering factor. 
The panoramic video and VR communities have extensively compared equirectangular, cubemap, and other spherical projection formats as rendering techniques, evaluating their effects on image quality, pixel density, and compression efficiency~\cite{hussain_evaluation_2021, shafi_360-degree_2020}.
Some work has examined how projection distortions affect perception in panoramic imagery~\cite{corbillon_viewport-adaptive_2017}. 
However, these studies treat projection as a display fidelity problem rather than a factor shaping communication between collaborators.

In collaborative settings, projection takes on additional significance because it defines the shared spatial frame of reference for the collaborating users. 
Despite the central role of spatial referencing in collaboration, the influence of projection strategy on cross-platform communication remains largely unexplored. 
Our work addresses this gap by systematically comparing four projection techniques across three layout curvatures to examine how projection choice shapes spatial referencing in cross-reality settings.
\section{\textsc{CrossAtlas}}
\label{system}
We designed and developed \textsc{CrossAtlas}, a synchronized cross-reality collaboration platform for studying how bidirectional projection techniques shape spatial understanding between a VR collaborator and a PC collaborator.
In \textsc{CrossAtlas}, the VR user interacts with a surrounding 3D point layout, while the PC user views a 2D representation of that same layout. 
Both users operate on the same underlying spatial environment across different geometric representations while communicating with each other, with the system maintaining synchronization between the 3D environment in VR and the 2D view on the PC so that selections and references in one representation can be consistently interpreted in the other.
The system is implemented as a lightweight web application, allowing both clients to run in the browser without platform-specific installation.
In the following subsections, we describe the key design elements and implementation details of \textsc{CrossAtlas}.

\begin{figure}
  \centering
  \includegraphics[width=\linewidth]{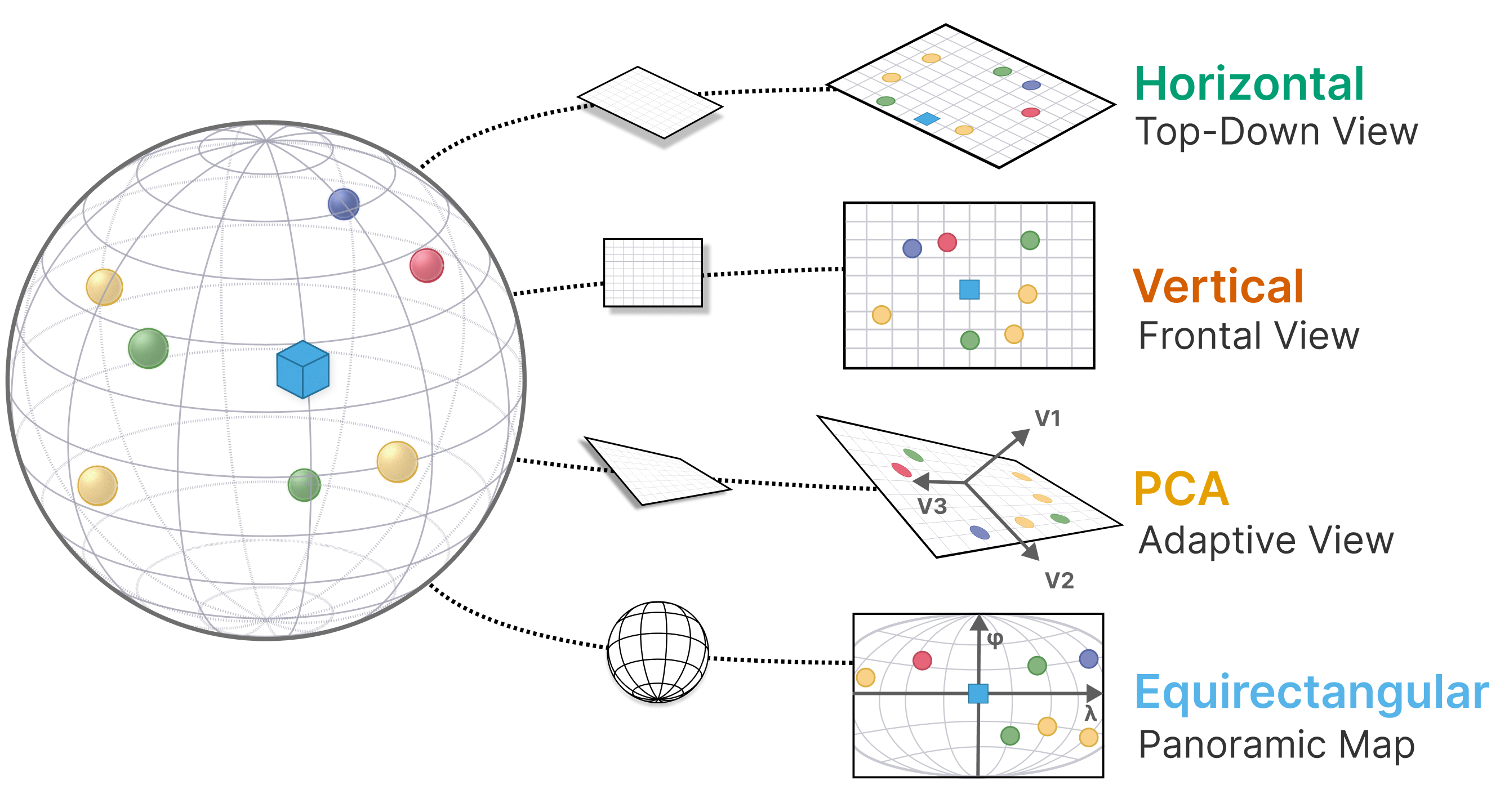}
  \caption[Projection techniques in \textsc{CrossAtlas}]{
    Illustration of the four projection techniques implemented in \textsc{CrossAtlas} to transform the VR user's 3D environment into 2D representations for the PC collaborator. The first three are \textbf{planar projections}: (A) \textcolor{horizontal}{\textit{\textbf{horizontal}}}, a top-down view that preserves lateral layout; (B) \textcolor{vertical}{\textit{\textbf{vertical}}}, a front-facing view that compresses depth; and (C) \textcolor{pca}{\textit{\textbf{PCA}}}, an adaptive view that maximizes variance in the projected point distribution. The fourth is a \textbf{spherical projection}: (D) \textcolor{equirectangular}{\textit{\textbf{equirectangular}}}, which provides a continuous 360\degree{} panoramic view of the surrounding space. A shared landmark (blue cube or square) is synchronized across all conditions and remains visible in both views.
  }
  \label{fig:projections}
  \vspace{-1em}
\end{figure}

\subsection{Bidirectional Projection Techniques}
Bidirectional projection is the core design dimension in \textsc{CrossAtlas} because it determines how 3D object coordinates in the shared spatial layout are transformed into 2D representations in real time. 
In the forward direction, each object's 3D position is mapped to a 2D coordinate in the PC view according to the active projection technique. 
For the inverse mapping, the system stores each object's depth from the VR user, allowing selections in the 2D view to be mapped back to the corresponding 3D object without additional depth inference.
This bidirectional correspondence underlies the full set of projection techniques implemented in the system.

To support systematic comparison, the system implements four projection techniques drawn from two projection families: \textbf{planar projection} and \textbf{spherical projection}.

\smallskip
\noindent\textbf{Planar Projection.}
In planar projection, each 3D point is mapped onto a plane by projecting it along a perpendicular direction, preserving the arrangement of points as seen from a fixed viewpoint. 
Drawing from existing approaches, \textsc{CrossAtlas} implements three variants: \textit{horizontal} and \textit{vertical} planes aligned with the VR user's canonical reference frames, and an adaptive \textit{PCA} plane that follows prior cross-reality systems and maximizes variance in the projected point distribution~\cite{borowski_spatialstrates_2025}. 
We excluded interactive projection-plane control because it would introduce individual view-control strategies and navigation behaviors, making it more difficult to isolate the effect of projection technique.

\smallskip
\noindent\raisebox{-0.2em}{\includegraphics[height=1em]{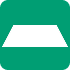}}\hspace{0.3em}\textcolor{horizontal}{\textit{\textbf{horizontal.}}}
Points are projected onto a horizontal plane positioned beneath the VR user (i.e., the floor)~(\autoref{fig:projections}\figpart{A}).
This is equivalent to viewing the point layout from directly above, producing a top-down representation. 
Horizontal relationships between points (such as left--right and front--back) are preserved, but all vertical structure is collapsed.

\smallskip
\noindent\raisebox{-0.2em}{\includegraphics[height=1em]{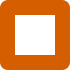}}\hspace{0.3em}\textcolor{vertical}{\textit{\textbf{vertical.}}}
Points are projected onto a vertical plane facing the VR user (i.e., a wall directly in front of them)~(\autoref{fig:projections}\figpart{B}). This produces a front-facing representation that preserves left--right and up--down relationships as seen from the user's forward-facing perspective, but compresses depth: points at different distances from the user that share the same lateral and vertical position overlap.

\smallskip
\noindent\raisebox{-0.2em}{\includegraphics[height=1em]{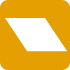}}\hspace{0.3em}\textcolor{pca}{\textit{\textbf{PCA.}}}
Points are projected onto a plane whose orientation is determined via PCA, which selects the plane that maximizes the variance of the projected point distribution~(\autoref{fig:projections}\figpart{C}).
PCA computes the two principal axes of greatest spread in the 3D point cloud and uses these to define the projection plane. 
The plane is centered on the VR user's initial position at the start of the task and normalized in scale to remain consistent with the other planar projections.
To ensure a stable and interpretable output, one basis vector is aligned with the vertical gradient direction (establishing a consistent ``up'' in the projected view), and the second is chosen to avoid producing a mirrored image relative to the VR user's perspective. 
When the PCA plane is approximately horizontal, the system defaults to the horizontal planar orientation.
This adaptive approach maximizes variance in the projected coordinates, helping preserve spatial group structure and separation in the 2D view, though the resulting projection plane may be less intuitive than the predefined canonical planes.

\smallskip
\noindent\textbf{Spherical Projection.}
In contrast to planar projections, which depict the scene from a single viewing direction, spherical projections from cartography represent a full surrounding environment within a single continuous 2D view. 
These techniques transform spherical coordinates onto flat surfaces, enabling complete 360\degree{} coverage of the environment. 
Many spherical projections exist (e.g., Mercator, Mollweide, equirectangular), each introducing different trade-offs in area, shape, and distance distortion. 
We chose to implement \textit{equirectangular} projection in \textsc{CrossAtlas} because it provides a simple, screen-friendly mapping from spherical coordinates to a rectangular 2D view.
While prior work has shown that the choice among spherical projections can affect geographic task performance, its effect on layout-based tasks may be less pronounced and task-dependent~\cite{chen_gansda_2022}.

\smallskip
\noindent\raisebox{-0.2em}{\includegraphics[height=1em]{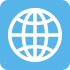}}\hspace{0.3em}\textcolor{equirectangular}{\textit{\textbf{equirectangular.}}}
Points are first projected onto a unit sphere centered on the VR user, mapping each point to its angular position. 
The resulting spherical coordinates (longitude $\lambda$ and latitude $\phi$) are mapped linearly to the horizontal and vertical axes of a 2D rectangle, respectively~(\autoref{fig:projections}\figpart{D}). 
As a result, the horizontal axis represents the full 360\degree{} sweep around the VR user, while the vertical axis encodes elevation from below to above the user.
This mapping produces a continuous panoramic representation of the environment, enabling the PC user to observe every point in the layout within a single unified view, including points behind the VR user.
Spatial relationships are expressed in terms of angular position around the VR user's body-centered frame, supporting references such as objects being located to the left, right, or behind the user.
The trade-off is systematic geometric distortion: areas near the poles (directly above and below the user) are stretched horizontally, and distances between points are not uniformly represented across the image. 

\subsection{Cross-View Synchronization and Shared Reference}
\label{system_sync}
To support collaboration across heterogeneous views, \textsc{CrossAtlas} synchronizes all task-relevant state between the VR and PC clients in real time. 
When either user selects an object, the event is propagated immediately to the other client. 
The system also synchronizes point visibility, selection outcomes, visual feedback, task progression, target assignment, and condition transitions through a shared state model subscribed to by both clients.

Synchronization is implemented with Yjs~\cite{yjs}, which uses conflict-free replicated data types (CRDTs) to maintain a consistent distributed state without centralized conflict resolution for each interaction. 
The shared document is communicated through a WebSocket provider over a TLS-encrypted connection (WSS).
In deployment, the system showed no noticeable delay during testing or the user study.

\textsc{CrossAtlas} also includes a virtual landmark as a synchronized shared reference point visible in both the 3D and 2D views. 
The landmark is visually distinct from ordinary objects and is typically initialized near the center of each user's view. 
Its purpose is to provide a stable cross-view anchor for spatial reference, enabling descriptions such as ``left of the landmark'' without requiring collaborators to first establish a common coordinate system. 
This design simulates common collaborative practices in which participants rely on shared reference objects (such as title cards or prompts in digital whiteboarding systems) to organize and navigate an otherwise open workspace. 
It is also motivated by prior work showing that landmarks support spatial memory and orientation in immersive visualization environments~\cite{liu_investigating_2024}. 
Accordingly, the landmark functions as an intentionally shared object-relative anchor that remains available across all conditions.

\subsection{Configurable Spatial Layouts}
\label{system_layout}
Beyond projection, \textsc{CrossAtlas} treats the geometry of the VR workspace itself as configurable.
Prior immersive analytics research has shown that users often organize information in 3D space using recurring spatial patterns, such as planar surfaces in front of them, curved arcs around the body, or surrounding spherical layouts that support visibility in large information spaces~\cite{batch_there_2020, lee_shared_2021}.
These layouts emerge as users balance spatial memory, reachability, and perceptual coverage when working with large datasets in immersive environments~\cite{lisle_sensemaking_2021, luo_where_2022}.
Reflecting these observed patterns, \textsc{CrossAtlas} places target objects around the egocentric VR user within a room-scale environment and supports variation in layout curvature. 
Layouts may range from a flat, front-facing arrangement that confines content to a narrow region in front of the user to increasingly curved and ultimately fully surrounding configurations that wrap content around the user's body-centered space~(\autoref{fig:layout}).
As content wraps further around the user, interpretation from a PC view becomes increasingly dependent on how the active projection preserves, compresses, or distorts those regions of space.

\textsc{CrossAtlas} further controls spatial concentration by parameterizing the cumulative angular distance between targets as measured from the VR user's headset position. This allows layout geometry to be tuned relative to the user's body-centered frame rather than treated as an arbitrary set of coordinates.
Separating layout geometry from projection techniques allows the system to vary the structure of the VR space independently of the transformation used to depict it on the PC.

\subsection{Technical Implementation}
\textsc{CrossAtlas} is implemented as a browser-based web platform with separate rendering layers for VR and PC clients. The VR client is built with Babylon.js~\cite{babylonjs} and extended with Anu.js~\cite{anujs}.
Running within WebXR~\cite{webxr} makes the VR client largely hardware-agnostic for headsets with WebXR-capable browsers. 
The PC client is implemented in React and renders the active 2D representation in a standard browser window. 
Together, these components provide a lightweight cross-reality platform that integrates bidirectional projection, cross-view synchronization, and layout configuration within a unified web-based pipeline.
\section{User Study}
We conducted a controlled lab study to evaluate how projection techniques in \textsc{CrossAtlas} influence collaborative spatial referencing between a VR user and a PC user under VR layouts of varying curvature. 
Specifically, we examined how projection type and layout curvature affect a dyad's ability to achieve \textit{spatial congruency} and accurately identify target objects across heterogeneous views. 
To do so, collaborating dyads completed structured spatial referencing tasks under a partially crossed experimental design. 

\begin{figure}
  \centering
  \includegraphics[width=\linewidth]{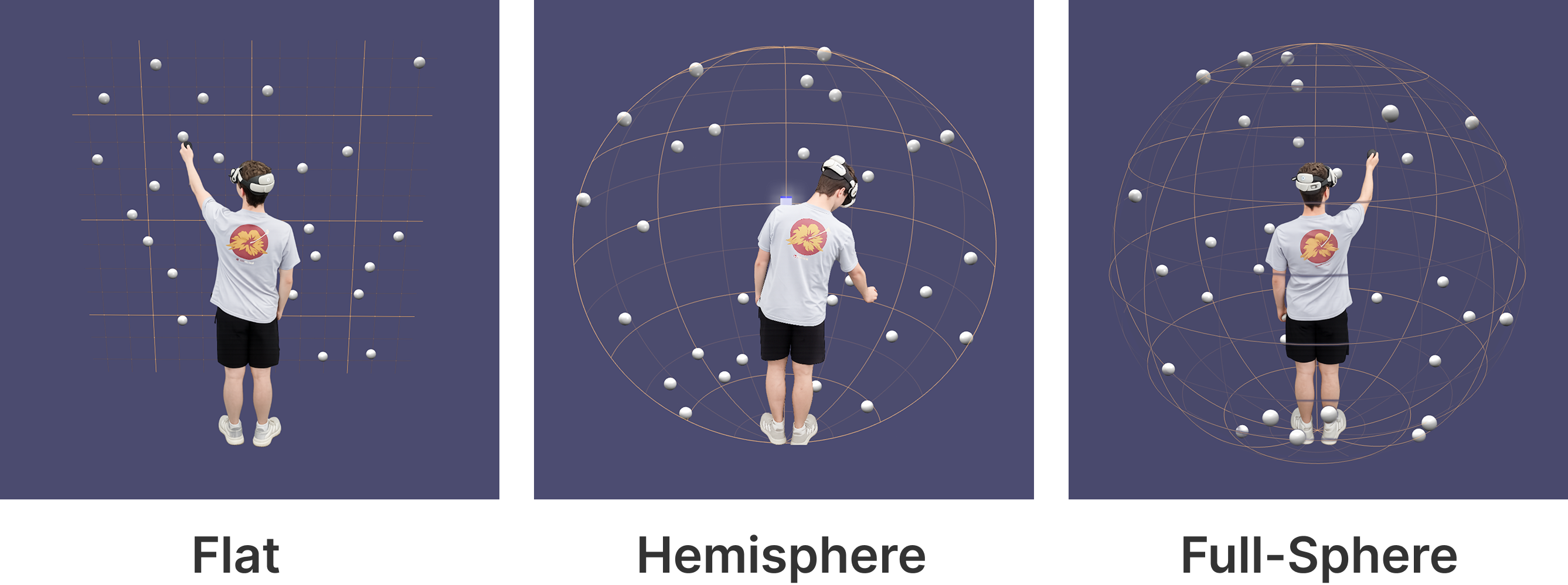}
  \caption[Layout curvature conditions]{
    The three layout curvature conditions used in the user study: \textbf{Flat}, \textbf{Hemisphere}, and \textbf{Full-Sphere}. As curvature increases, targets extend from a front-facing planar region to a fully surrounding arrangement around the VR user.
  }
  \label{fig:layout}
  \vspace{-1.5em}
\end{figure}

\subsection{Study Design and Hypotheses}
We employed a partially within-subjects $4 \times 3$ factorial design, manipulating Projection Techniques (\textit{horizontal}, 
\textit{vertical}, 
\textit{PCA},
\textit{equirectangular}) 
and Layout Curvature (3 levels: 
\textbf{Flat}, 
\textbf{Hemisphere}, 
\textbf{Full-Sphere}). 
Due to time constraints, each dyad completed all four projection conditions under two of the three layout curvature conditions, yielding eight conditions per dyad. 
Layout assignments were counterbalanced across dyads so that all three curvature levels were equally represented across the dataset.
Using \textsc{CrossAtlas}'s configurable spatial layout generation~(\autoref{system_layout}), we instantiated the three layout curvature conditions at varying point densities~(\autoref{fig:layout}) while maintaining comparable task difficulty across conditions.
To isolate projection geometry, we used abstract targets, static layouts, co-located dyads, minimal awareness cues beyond task feedback and a shared landmark, and fixed device roles (one participant used VR and the other used a PC throughout); we return to these controls and their implications in~\autoref{limitations}.

Each layout curvature condition constituted one task block.
Within each block, participants completed three tasks for each of the four projection techniques in succession. 
Projection order was counterbalanced across dyads using a 4-condition Williams Latin square to balance position and first-order carryover effects. 
To standardize task difficulty across conditions, the cumulative angular distance between target points (as measured from the VR user's headset position at the start) was controlled within each task so that the target separations across its trials summed to the same total.

Based on a pilot study with a dyad, findings from related work, and \textsc{CrossAtlas}'s design rationales~(\autoref{system}), we preregistered the following hypotheses on OSF\footnote{Preregistration is available at \url{https://osf.io/9q6mz/overview}.}:

\begin{enumerate}[
    topsep=4pt,
    itemsep=3pt,
    parsep=0pt,
    partopsep=0pt,
    leftmargin=*,
    labelwidth=2em,
    labelsep=0.5em,
    align=left,
    label=\textbf{H\arabic*},
    ref=H\arabic*
]
\item\label{h1}
\textbf{Projection.}
\textit{Equirectangular} would outperform the three planar projections (\textit{PCA}, \textit{horizontal}, and \textit{vertical}) in accuracy, completion time, and subjective ratings.

\item\label{h2}
\textbf{Layout Curvature.}
Increasing layout curvature would reduce task performance, with \textbf{Flat} layouts yielding the best performance and \textbf{Full-Sphere} layouts yielding the worst.

\item\label{h3}
\textbf{Projection\hspace{0.08em}$\times$\hspace{0.12em}Layout Interaction.}
The advantage of \textit{equirectangular} over the planar projections would increase with layout curvature, with the largest gap in \textbf{Full-Sphere} layouts.

\item\label{h4}
\textbf{Projection\hspace{0.08em}$\times$\hspace{0.12em}Task Complexity Interaction.}
The advantage of \textit{equirectangular} would increase with task complexity, with the smallest gap in Task~1 and the largest in Task~3.
\end{enumerate}

\subsection{Tasks}
For each projection condition, dyads completed seven trials: four trials of Task~1, two trials of Task~2, and one trial of Task~3.
The three tasks modeled common forms of collaborative spatial referencing with progressively greater coordination demands. 
Task~1 represented simple one-way reference, such as a VR user asking a PC collaborator to inspect ``the window behind me.'' 
Task~2 required both collaborators to describe and interpret different locations. Task~3 required them to establish common ground over time by using previously identified locations as anchors for subsequent references.

Across all tasks, objects appeared as spheres in VR and circles in the PC projection, representing the same underlying points in the shared layout.
Task-relevant targets were marked in green (or the assigned pair color in Task~3) only for the participant who could see them, while incorrect selections caused the corresponding object to flash red.
These visual cues were synchronized in real time to provide consistent feedback across both platforms.
A shared landmark (a blue cube in VR and a blue square on PC), as described in~\autoref{system_sync}, remained visible to both users and served as an initial anchor for spatial reference.

\smallskip
\noindent
\textbf{Task~1: Simple Spatial Reference.}
One participant was shown a target object visible only to them and verbally described its location so that their partner could select it.
Descriptions could refer to the layout geometry (e.g., ``\textit{bottom right},'' ``\textit{far left}'') or the shared landmark (e.g., ``\textit{two spheres left of the landmark}'').
Participants alternated between giving and following instructions across successive trials. Inspired by prior work on spatial reference in asymmetric-view collaboration, this task served as a baseline for examining how one-way references were produced and interpreted across views~\cite{johnson_you_2021,muller_remote_2017}.

\smallskip
\noindent
\textbf{Task~2: Mutual Spatial Reference.} 
Each participant was shown a different target object that was not visible to their partner.
The VR user's target was additionally marked with a glow to distinguish it from the PC user's target of the same color.
Both participants exchanged spatial information to help their partner locate one target while gathering the information needed to locate their own. 
A trial was successful only when both selections were correct. 
Compared with Task~1, this task required both participants to act as information providers and seekers, emphasizing mutual awareness, turn-taking coordination, and perspective-taking across views~\cite{saffo_through_2023, numan_exploring_2022}.

\smallskip
\noindent
\textbf{Task~3: Color-Pair Matching.} 
Participants identified a sequence of colored target pairs, with red excluded because it indicated incorrect selections.
During each turn, the system revealed a target color corresponding to one pair, and both participants exchanged spatial references while locating their respective targets.
When both targets in a pair were correctly selected, the pair remained permanently visible to both participants, creating additional shared anchors for subsequent turns within the same trial. 
As more pairs were identified, participants progressively constructed and reused a shared set of reference points.
This task therefore examined whether a projection continued to support collaboration as the shared history of the interaction increasingly structured the workspace~\cite{dourish_awareness_1992}.

\begin{figure}
  \centering
  \includegraphics[width=\linewidth]{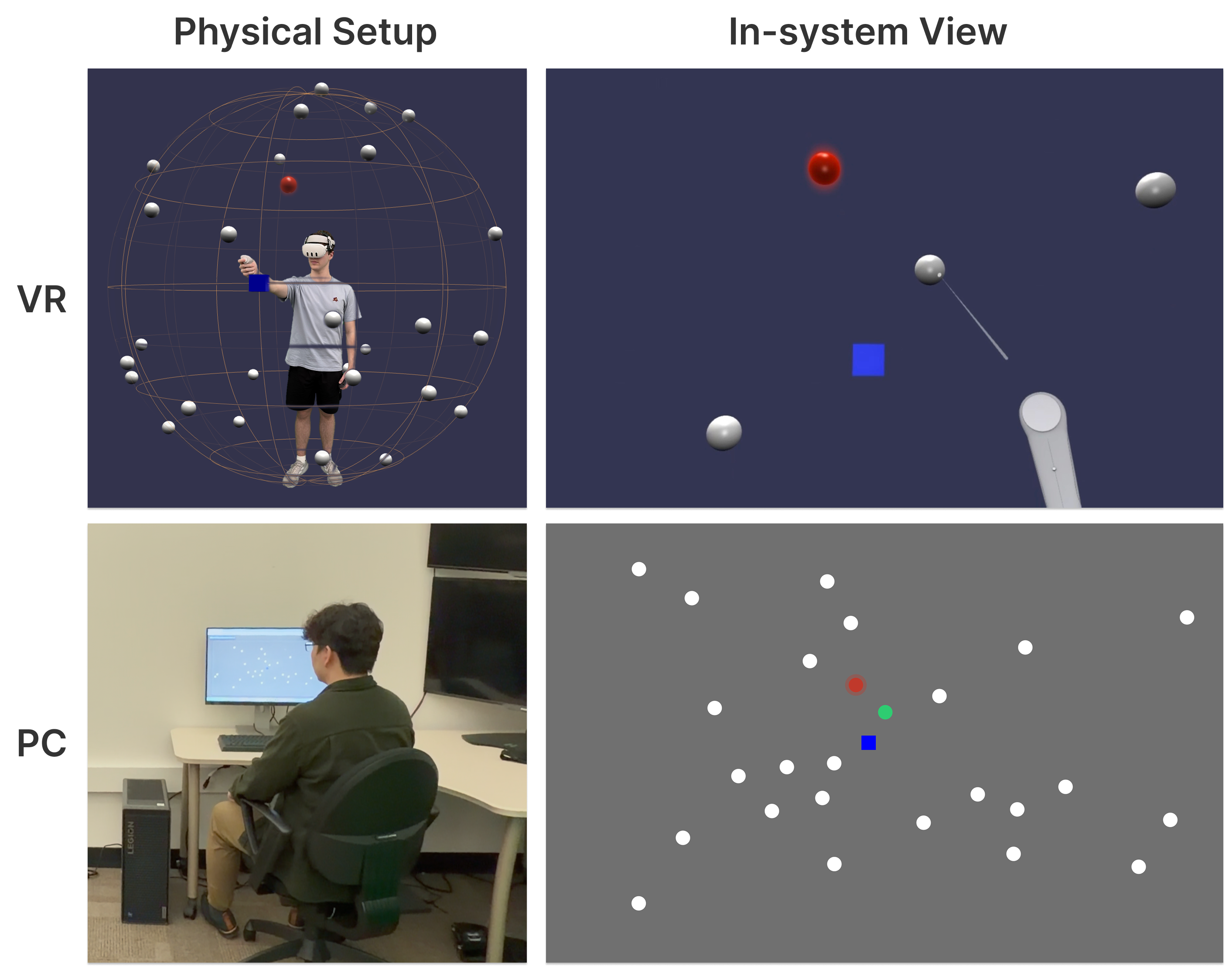}
  \caption[Experimental setup]{
    Experimental setup in the user study. The left column shows the physical setup, and the right column shows the corresponding in-system views for the VR and PC collaborators.
  }
  \label{fig:setup}
  \vspace{-1.5em}
\end{figure}

\subsection{Procedure}
Each session lasted approximately 90 minutes and followed a fixed sequence.

\smallskip
\noindent
\textbf{Orientation and Training (20 minutes).}
Participants first provided informed consent and completed a brief demographic survey and a computerized version of the Perspective-Taking/Spatial Orientation Test (PTSOT)~\cite{kozhevnikov_dissociation_2001, hegarty_dissociation_2004, friedman_computerized_2020} to measure baseline spatial perspective-taking ability.
Participants were then randomly assigned to either the VR or PC role. 
The experimenter introduced the corresponding interfaces and guided the dyad through practice trials, focusing on target selection and verbal spatial referencing.

\smallskip
\noindent 
\textbf{Task Blocks (70 minutes).} 
Each dyad completed two task blocks corresponding to their assigned layout curvature conditions. 
Each block included approximately 30 minutes of task execution followed by a 5-minute post-block assessment, in which participants evaluated and ranked the projection techniques using workload and Spatial Experience questionnaires.
Participants were free to develop their own communication strategies, with the only restriction being that targets from previous trials could not be used as spatial anchors, as this would trivialize the task.

\subsection{Participants and Apparatus}
We recruited 24 dyads (48 participants, aged 18--31, M = 24.0, SD = 3.24; 27 male, 21 female). 
Most participants (87.5\%) enrolled in the study with a partner they already knew. The remaining participants were paired by the experimenters with a previously unacquainted partner.
VR experience varied: 7 had no prior experience, 25 had tried VR 1--2 times, 10 used it occasionally, and 6 were frequent users. 
On the PTSOT, participants showed a mean angular error of 24.03\degree{} (SD = 20.28\degree{}, range = 6.59\degree{}--94.79\degree{}), where lower scores indicate better perspective-taking ability.
Relative to prior published adult samples~\cite{karamazovova_spatial_2025, gunalp_spatial_2019}, this suggests generally typical to slightly above-average spatial perspective-taking performance at baseline.
All participants reported normal or corrected-to-normal vision and no color blindness. Each participant received a \$20 Amazon gift card as compensation.

The experiment was conducted in a $10 \times 10$ ft lab area, with both participants co-located in the same room~(\autoref{fig:setup}). 
The VR participant used a Meta Quest 3 headset with sufficient space for physical movement, while the PC participant completed the study on a Windows laptop using a browser-based interface in Chrome.
The WebXR application and Yjs backend were hosted separately on a MacBook Pro. The headset, PC, and server were connected through a dedicated Wi-Fi 7 router.

\subsection{Data Collection}
We collected performance measures through system logs, including completion time, selections, correctness, attempts, and incorrect selections. 
Subjective measures included NASA-TLX and a custom Spatial Experience questionnaire using a 7-point Likert scale. 
The questionnaire comprised four self-location items adapted from the Spatial Presence Experience Scale (SPES)~\cite{hartmann_spatial_2016} and six perceived \textit{spatial congruency} items informed by prior work~\cite{saffo_through_2023,wang_mrtransformer_2024,borowski_spatialstrates_2025}, which assessed cross-view alignment, ease of translating spatial references, and shared understanding of object locations. 
After confirming internal consistency, all ten items were averaged into a composite Spatial Experience score, with higher scores indicating a more coherent spatial experience. 
The full questionnaire is provided in the supplementary materials.

\subsection{Statistical Analysis}

We analyzed objective performance measures and subjective ratings using mixed-effects models implemented in R~\cite{bates2015fitting,bolker2009generalized,lenth2016least}. Mixed-effects modeling was chosen because it accommodates repeated observations from the same participants and handles unbalanced experimental designs while avoiding the sphericity assumption required by repeated-measures ANOVA~\cite{field2012discovering}. Completion time was positively skewed and therefore log-transformed prior to analysis. The transformed completion time was analyzed using a linear mixed-effects model with layout, projection, task, and their interactions as fixed effects and dyad as a random intercept. Because the error variable (\texttt{sum\_incorrect}) consisted of count data and exhibited overdispersion, it was analyzed using a generalized linear mixed-effects model with a negative binomial distribution and log link~\cite{hilbe2011negative}, again including layout, projection, task, and their interactions as fixed effects and dyad as a random intercept.

Subjective ratings were analyzed using the same modeling framework. The six NASA-TLX dimensions were rescaled to 1--7, averaged, and reverse-coded so that higher scores indicate lower workload; the ten Spatial Experience items were averaged on their original scale. Each composite, as well as each individual NASA-TLX dimension and Spatial Experience item, was analyzed separately using a linear mixed-effects model with projection, layout, and platform as fixed effects, participant as a random intercept, and platform as a between-subjects factor. Fixed effects were evaluated using Type III ANOVA with Satterthwaite's approximation, whereas the negative binomial model used joint Wald chi-square tests. Post hoc analyses were conducted using estimated marginal means (EMMs) with 95\% confidence intervals (CIs) and Tukey-adjusted pairwise comparisons~\cite{lenth2016least}. Model assumptions were assessed using diagnostic plots, including histograms, density plots, and Q--Q plots for log-transformed completion time, as well as dispersion diagnostics for the error counts. Statistical significance is reported as $p < .05$ (*), $p < .01$ (**), and $p < .001$ (***); full statistical outputs are provided in the supplementary materials.
\section{Results}

Both completion time and error count showed significant effects of layout, projection, and task, as well as a layout $\times$ projection interaction (all $p < .001$). Composite subjective ratings showed a similar projection-dependent pattern: both NASA-TLX and Spatial Experience scores showed significant effects of layout and projection, as well as a layout $\times$ projection interaction (all $p < .001$). Below, pairwise significance is denoted as $*$ ($p<.05$), $**$ ($p<.01$), and $***$ ($p<.001$).

\begin{figure}
    \centering
    \includegraphics[width=\linewidth]{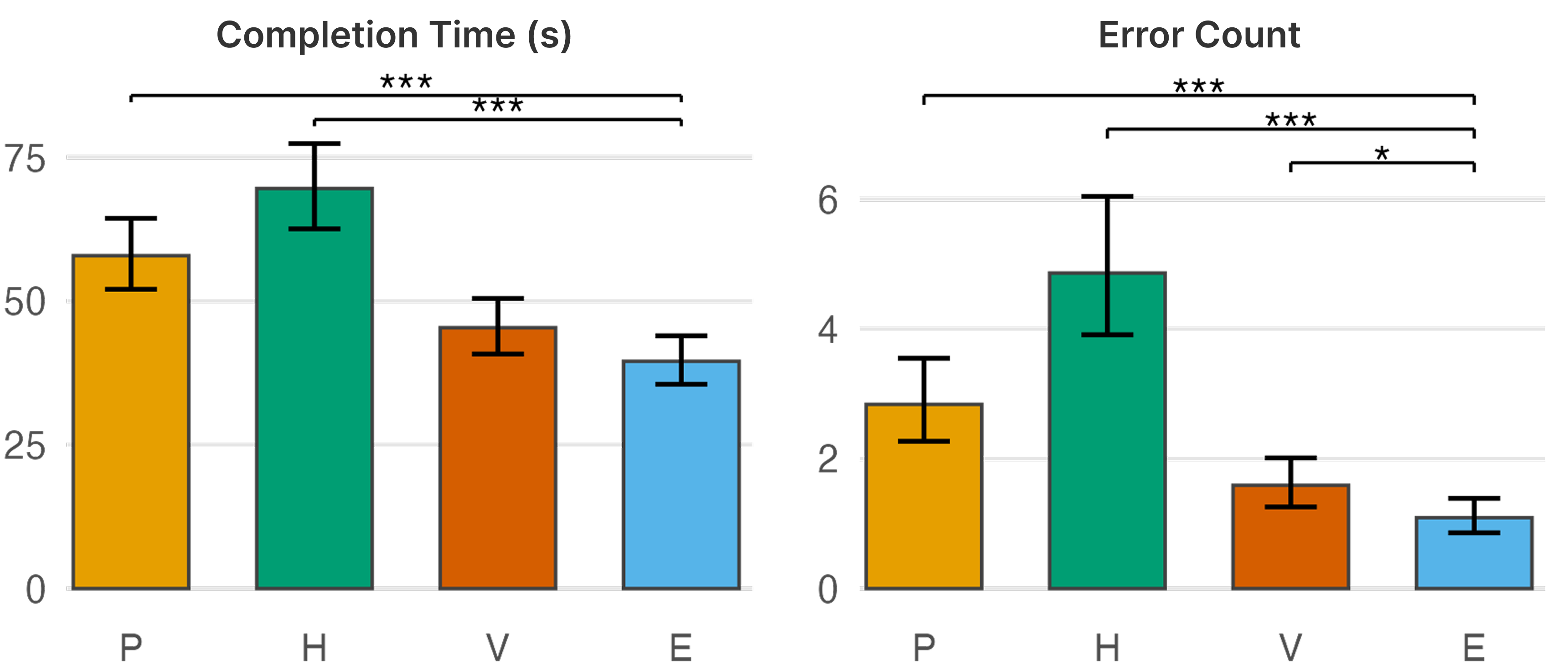}
    \caption[Projection effect on completion time and error count]{
    Model-estimated marginal means of projection effect on completion time (left) and error count (right), averaged across layouts and tasks. 
    Error bars indicate 95\% confidence intervals (CIs); brackets indicate significant pairwise differences (* $p < .05$, ** $p < .01$, *** $p < .001$).
    Projections include \textcolor{pca}{\textit{\textbf{PCA}}} (P), \textcolor{horizontal}{\textit{\textbf{horizontal}}} (H), \textcolor{vertical}{\textit{\textbf{vertical}}} (V), and \textcolor{equirectangular}{\textit{\textbf{equirectangular}}} (E).
    }
    \label{fig:projection-effect}
    \vspace{-1em}
\end{figure}

\subsection{Projection Effect}

\smallskip
\noindent\textbf{Completion Time.}
Collapsed across layouts and tasks, \textit{equirectangular} yielded the shortest completion time (39.51\,s, CI=8.42\,s), followed by \textit{vertical} (45.36\,s, CI=9.67\,s), \textit{PCA} (57.89\,s, CI=12.34\,s), and \textit{horizontal} (69.57\,s, CI=14.83\,s)~(\autoref{fig:projection-effect}).
\textit{Equirectangular} was significantly faster than \textit{horizontal} and \textit{PCA} (both $***$), but not \textit{vertical}. The overall projection effect was thus driven by slower performance under both \textit{horizontal} and \textit{PCA}, with \textit{equirectangular} and \textit{vertical} forming the faster group.

\smallskip
\noindent\textbf{Error Count.}
\textit{Equirectangular} also produced the fewest errors overall (1.09, CI=0.53), followed by \textit{vertical} (1.59, CI=0.75), \textit{PCA} (2.84, CI=1.28), and \textit{horizontal} (4.86, CI=2.13). It yielded significantly fewer errors than \textit{horizontal} ($***$), \textit{PCA} ($***$), and \textit{vertical} ($*$). Overall, these results support a clear projection ranking in accuracy, with \textit{horizontal} again performing worst.

\smallskip
\noindent\textbf{Subjective Ratings.}
Across layouts and platforms, \textit{equirectangular} received the highest ratings on both NASA-TLX (5.53, CI=0.25) and Spatial Experience (5.26, CI=0.27)~(\autoref{fig:result_ratings}).
For NASA-TLX, it scored higher than \textit{horizontal} (3.55, CI=0.25) and \textit{PCA} (4.59, CI=0.25) (both $***$), but did not differ from \textit{vertical} (5.18, CI=0.25).
For Spatial Experience, it outperformed \textit{horizontal} (3.07, CI=0.27) ($***$), \textit{PCA} (4.10, CI=0.27) ($***$), and \textit{vertical} (4.81, CI=0.27) ($*$).
Overall, subjective ratings showed a clear preference for \textit{equirectangular} over \textit{horizontal} and \textit{PCA}, with a weaker contrast relative to \textit{vertical}.

\begin{figure}
    \centering
    \vspace{0.5em}
    \includegraphics[width=\linewidth]{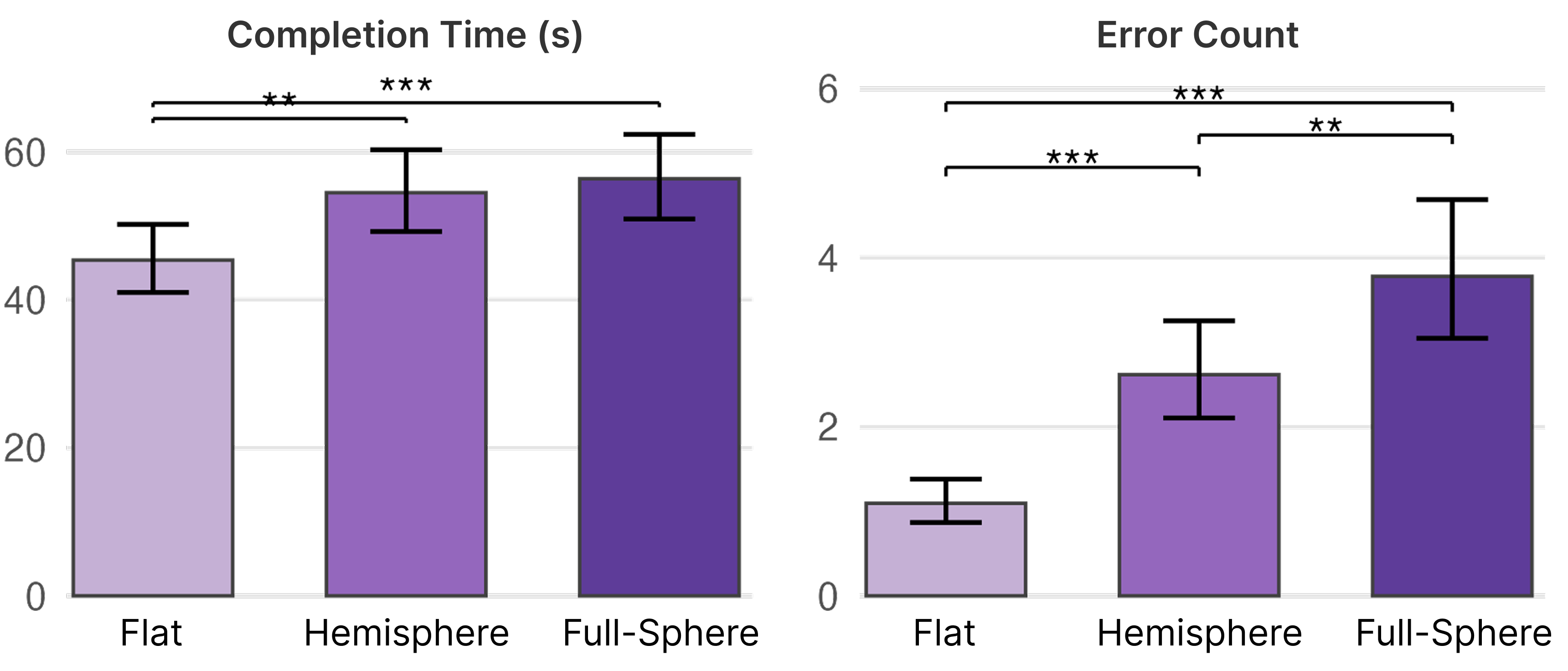}
    \caption[Layout effect on completion time and error count]{
    Model-estimated marginal means of layout effect on completion time (left) and error count (right), averaged across projections and tasks.
    Error bars indicate 95\% CIs; brackets indicate significant pairwise differences (* $p < .05$, ** $p < .01$, *** $p < .001$).
    Layouts include \textbf{Flat}, \textbf{Hemisphere}, and \textbf{Full-Sphere}.
    }
    \label{fig:layout-effect}
    \vspace{-1.5em}
\end{figure}

\subsection{Layout Effect}

\smallskip
\noindent\textbf{Completion Time.}
Across projections and tasks, Flat yielded the shortest completion time (45.36\,s, CI=9.20\,s), compared with Hemisphere (54.47\,s, CI=11.05\,s) and Full-Sphere (56.35\,s, CI=11.43\,s)~(\autoref{fig:layout-effect}). Flat was significantly faster than both Hemisphere ($**$) and Full-Sphere ($***$), whereas the latter two did not differ. This indicates that increased curvature slowed performance overall, although the difference was concentrated between Flat and the curved layouts.

\smallskip
\noindent\textbf{Error Count.}
Flat also produced the fewest errors overall (1.09, CI=0.51), followed by Hemisphere (2.62, CI=1.15) and Full-Sphere (3.78, CI=1.64). All pairwise contrasts were significant: Flat yielded fewer errors than Hemisphere and Full-Sphere (both $***$), and Hemisphere yielded fewer errors than Full-Sphere ($**$). The accuracy results showed a consistent order: Flat $<$ Hemisphere $<$ Full-Sphere.

\begin{figure*}
  \centering
  \includegraphics[width=\linewidth]{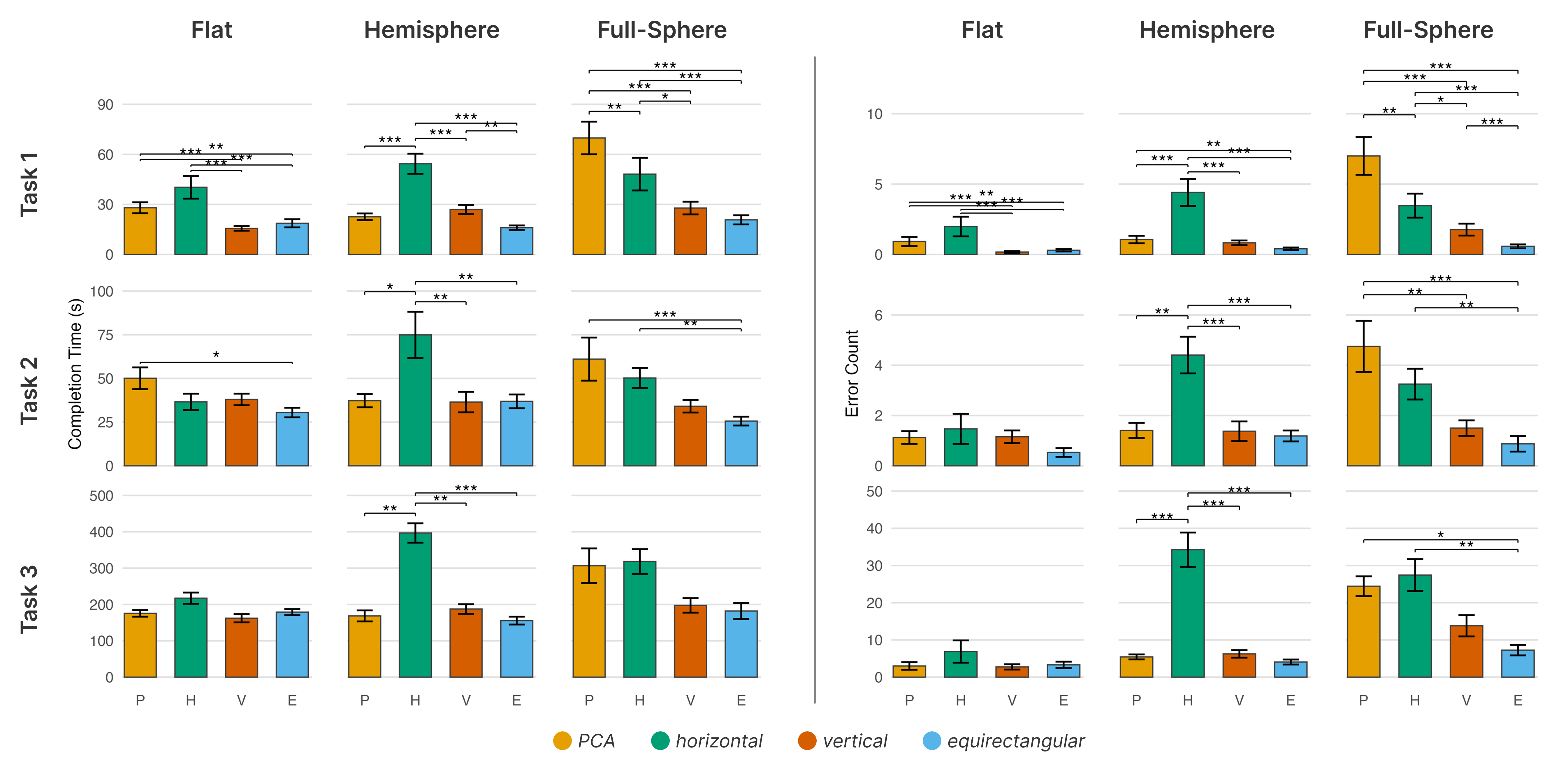}
  \caption[Projection performance across layouts and tasks]{
    Completion time (left) and error count (right) across four projections for three layout conditions (\textbf{Flat}, \textbf{Hemisphere}, and \textbf{Full-Sphere}) and three tasks. Error bars indicate 95\% CIs; brackets indicate significant pairwise differences (* $p < .05$, ** $p < .01$, *** $p < .001$).
  }
  \label{fig:results_projections}
  \vspace{-1.5em}
\end{figure*}

\smallskip
\noindent\textbf{Subjective Ratings.}
Across projections and platforms, Flat received the highest ratings overall~(\autoref{fig:result_ratings}). For NASA-TLX, Flat (5.39, CI=0.24) scored higher than both Hemisphere (4.56, CI=0.24) and Full-Sphere (4.19, CI=0.24) (both $***$), and Hemisphere also exceeded Full-Sphere ($*$), yielding a clear ordering of Flat $>$ Hemisphere $>$ Full-Sphere. Spatial Experience ratings showed the same overall direction: Flat (4.71, CI=0.26) exceeded both Hemisphere (4.21, CI=0.26) ($**$) and Full-Sphere (4.00, CI=0.26) ($***$), whereas the difference between the two curved layouts was not significant.

\subsection{\texorpdfstring{Projection\hspace{0.1em}$\times$\hspace{0.15em}Layout Interaction}{Projection x Layout Interaction}}

\smallskip
\noindent\textbf{Completion Time.}
Projection effects on completion time depended on both task and layout~(\autoref{fig:results_projections}). 
In Task~1 under the Flat layout, \textit{vertical} (15.62\,s, CI=2.79\,s) and \textit{equirectangular} (18.68\,s, CI=4.85\,s) did not differ significantly, and both were significantly faster than \textit{horizontal} (40.26\,s, CI=13.60\,s) ($***$ for both comparisons). 
\textit{Equirectangular} was also faster than \textit{PCA} (28.00\,s, CI=6.56\,s) ($**$), and \textit{vertical} was faster than \textit{PCA} ($***$). 
In the Hemisphere layout, \textit{equirectangular} (16.06\,s, CI=2.69\,s) was significantly faster than \textit{horizontal} (54.38\,s, CI=12.03\,s) ($***$) and \textit{vertical} (26.96\,s, CI=5.36\,s) ($**$), while \textit{horizontal} was slower than both \textit{PCA} (22.61\,s, CI=3.97\,s) and \textit{vertical} (both $***$). 
For Full-Sphere, \textit{equirectangular} (20.76\,s, CI=5.48\,s) was faster than \textit{horizontal} (48.14\,s, CI=19.53\,s) ($***$) and \textit{PCA} (69.82\,s, CI=19.54\,s) ($***$); \textit{horizontal} was faster than \textit{PCA} ($**$) but slower than \textit{vertical} (27.82\,s, CI=7.61\,s) ($*$), and \textit{vertical} was faster than \textit{PCA} ($***$).

In Task~2, fewer projection differences were observed. In the Flat layout, only \textit{equirectangular} (30.49\,s, CI=5.58\,s) was faster than \textit{PCA} (50.10\,s, CI=12.68\,s) ($*$), while \textit{horizontal} (36.61\,s, CI=9.51\,s) and \textit{vertical} (37.98\,s, CI=6.74\,s) did not differ significantly from the other projections.
In the Hemisphere layout, \textit{horizontal} (74.94\,s, CI=26.93\,s) was slower than \textit{equirectangular} (36.89\,s, CI=8.03\,s) ($**$), \textit{PCA} (37.28\,s, CI=7.75\,s) ($*$), and \textit{vertical} (36.47\,s, CI=11.98\,s) ($**$). 
In the Full-Sphere layout, \textit{equirectangular} (25.58\,s, CI=5.14\,s) was faster than \textit{horizontal} (50.26\,s, CI=11.66\,s) ($**$) and \textit{PCA} (61.07\,s, CI=25.14\,s) ($***$), while the remaining contrasts were not significant.

In Task~3, no projection differences were significant in the Flat or Full-Sphere layouts. In the Hemisphere layout, however, \textit{horizontal} (396.60\,s, CI=56.77\,s) was significantly slower than \textit{equirectangular} (155.55\,s, CI=23.03\,s) ($***$), \textit{PCA} (168.49\,s, CI=32.40\,s) ($**$), and \textit{vertical} (187.38\,s, CI=28.33\,s) ($**$). Thus, across tasks, \textit{horizontal} consistently performed the worst.

\smallskip
\noindent\textbf{Error Count.}
Projection effects on error count largely mirrored the completion time results~(\autoref{fig:results_projections}). In Task~1 under the Flat layout, \textit{vertical} (0.17, CI=0.14) and \textit{equirectangular} (0.30, CI=0.17) did not differ significantly, but both yielded fewer errors than \textit{horizontal} (1.98, CI=1.39) ($***$ for both comparisons). \textit{Equirectangular} also yielded fewer errors than \textit{PCA} (0.92, CI=0.64) ($**$), and \textit{vertical} yielded fewer errors than \textit{PCA} ($***$). In the Hemisphere layout, \textit{horizontal} (4.41, CI=1.91) produced more errors than \textit{equirectangular} (0.41, CI=0.18), \textit{PCA} (1.06, CI=0.53), and \textit{vertical} (0.83, CI=0.33) (all $***$), and \textit{equirectangular} also outperformed \textit{PCA} ($**$). In the Full-Sphere layout, \textit{equirectangular} (0.58, CI=0.27) yielded fewer errors than \textit{horizontal} (3.47, CI=1.70), \textit{PCA} (7.00, CI=2.68), and \textit{vertical} (1.77, CI=0.85) (all $***$). \textit{Horizontal} also yielded fewer errors than \textit{PCA} ($**$), but more than \textit{vertical} ($*$), and \textit{vertical} outperformed \textit{PCA} ($***$).

In Task~2, no projection differences were significant in the Flat layout. In the Hemisphere layout, \textit{horizontal} (4.41, CI=1.49) again produced more errors than \textit{equirectangular} (1.19, CI=0.44) ($***$), \textit{PCA} (1.41, CI=0.61) ($**$), and \textit{vertical} (1.38, CI=0.80) ($***$). In the Full-Sphere layout, \textit{equirectangular} (0.88, CI=0.63) yielded fewer errors than \textit{horizontal} (3.25, CI=1.25) ($**$) and \textit{PCA} (4.75, CI=2.07) ($***$), while \textit{vertical} (1.50, CI=0.63) also outperformed \textit{PCA} ($**$).

In Task~3, no projection differences were significant in the Flat layout. In the Hemisphere layout, \textit{horizontal} (34.25, CI=9.83) produced more errors than \textit{equirectangular} (4.06, CI=1.44), \textit{PCA} (5.44, CI=1.43), and \textit{vertical} (6.25, CI=2.16) (all $***$). In the Full-Sphere layout, \textit{equirectangular} (7.25, CI=3.00) yielded fewer errors than \textit{horizontal} (27.44, CI=9.16) ($**$) and \textit{PCA} (24.44, CI=5.67) ($*$). Overall, \textit{horizontal} remained one of the least accurate projections in the curved layouts, whereas \textit{equirectangular} and \textit{vertical} generally maintained lower error counts.

\begin{figure*}
  \centering
  \includegraphics[width=\linewidth]{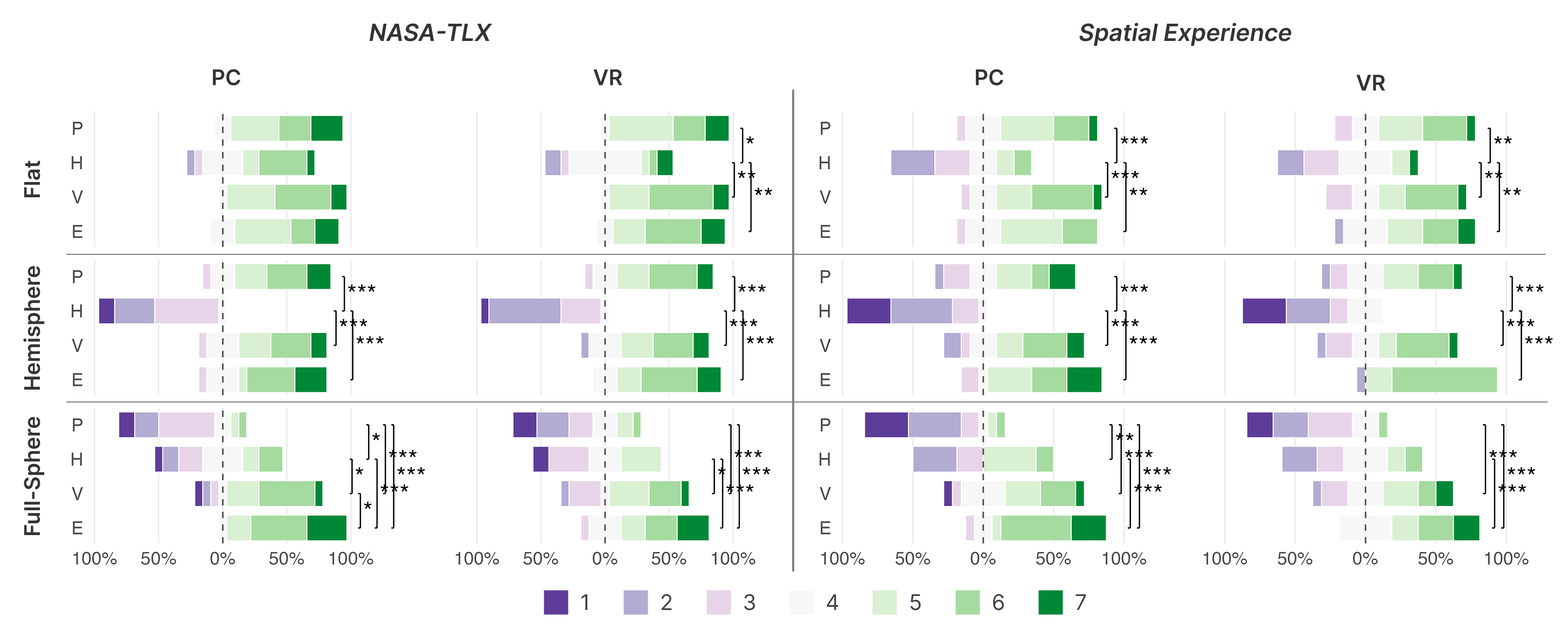}
  \caption[Subjective ratings across projection and layout conditions]{
      Distribution of subjective ratings for NASA-TLX (left) and Spatial Experience (right) across four projections---\textcolor{pca}{\textit{\textbf{PCA}}} (P), \textcolor{horizontal}{\textit{\textbf{horizontal}}} (H), \textcolor{vertical}{\textit{\textbf{vertical}}} (V), and \textcolor{equirectangular}{\textit{\textbf{equirectangular}}} (E)---for the three layout curvature conditions (\textbf{Flat}, \textbf{Hemisphere}, and \textbf{Full-Sphere}) and both platforms (PC and VR). NASA-TLX scores were reverse-coded so that higher values indicate lower workload. Stacked bars show the percentage distribution of Likert-scale responses. Dashed vertical lines mark the scale midpoint; brackets indicate significant pairwise differences (* $p < .05$, ** $p < .01$, *** $p < .001$).
  }
  \label{fig:result_ratings}
  \vspace{-1.5em}
\end{figure*}

\smallskip
\noindent\textbf{Subjective Ratings.}
This interaction pattern is also reflected in the aggregate ratings~(\autoref{fig:result_ratings}).
In both NASA-TLX and Spatial Experience ratings, \textit{horizontal} was generally the least preferred projection in the Flat (NASA-TLX: 4.63, CI=0.38; Spatial Experience: 3.57, CI=0.42) and Hemisphere layouts (NASA-TLX: 2.34, CI=0.38; Spatial Experience: 2.06, CI=0.42), with the strongest penalties appearing in the Hemisphere condition.
In the Flat layout, differences between \textit{horizontal} and the other projections were significant for Spatial Experience in both platforms ($**$ or $***$) and for NASA-TLX in VR only ($*$ or $**$). In the Hemisphere layout, \textit{horizontal} scored lower than all other projections on both ratings and platforms (all $***$).
In contrast, under the Full-Sphere layout, \textit{PCA} became the least preferred projection on both NASA-TLX (2.88, CI=0.38) and Spatial Experience (2.53, CI=0.42), whereas \textit{equirectangular} remained among the highest-rated options (NASA-TLX: 5.54, CI=0.38; Spatial Experience: 5.36, CI=0.42) and significantly outperformed \textit{PCA} on both ratings and platforms (all $***$).
Thus, subjective preference depended on layout: Flat and Hemisphere displays mainly penalized \textit{horizontal}, while the Full-Sphere layout most clearly favored \textit{equirectangular} over \textit{PCA}.

\subsection{\texorpdfstring{Projection\hspace{0.1em}$\times$\hspace{0.15em}Task Complexity Interaction}{Projection x Task Complexity Interaction}}

In general, the advantage of \textit{equirectangular} did not increase monotonically with task complexity. 
For completion time, \textit{equirectangular} was faster than both \textit{horizontal} and \textit{PCA} in Tasks~1 and~2, but in Task~3 only the difference from \textit{horizontal} remained significant; the difference from \textit{vertical} was significant only in Task~1. Error count showed the same pattern: \textit{equirectangular} outperformed \textit{horizontal} and \textit{PCA} in Tasks~1 and~2, but in Task~3 only the difference from \textit{horizontal} remained significant, and no difference from \textit{vertical} was significant.
\section{Discussion}

In this section, we discuss the findings from our user study in relation to our research questions. 
We structure the discussion around our hypotheses and study conditions, focusing on how projection techniques and layout curvature influenced different aspects of spatial referencing in collaborative tasks.

\smallskip
\noindent\textbf{Equirectangular projection provided a reliable advantage for collaborative spatial referencing tasks.}
Our results provided substantial support for our hypothesis~\textbf{\ref{h1}}: across tasks and layouts, \textit{equirectangular} yielded the shortest average completion time, the fewest errors, and the highest subjective ratings.
However, it did not consistently outperform every planar alternative. Among the planar projections, \textit{vertical} did not differ significantly from \textit{equirectangular} for overall completion time and NASA-TLX ratings, or in several task--layout conditions, whereas \textit{horizontal} was the weakest overall.

We believe that \textit{equirectangular} provides a continuous panoramic representation of the surrounding environment, allowing the PC user to view the full spatial context within a single unified frame. 
By preserving object locations in terms of angular position around the VR user's egocentric frame, it reduces ambiguity during cross-view interpretation and supports more stable spatial alignment between collaborators.
By contrast, planar projections impose stronger geometric constraints on the representation and can reduce \textit{spatial congruency} between collaborators when the spatial organization of the content is poorly aligned with the projection plane.
\textit{PCA}, although adaptive, may introduce a representation that is less immediately interpretable to users because its orientation is determined computationally rather than by an intuitive or familiar viewpoint. 
\textit{Vertical} performed more competitively, which may be because its frontal structure more closely matches the way users typically engage with screen-based interfaces, where content is presented directly in front of the viewer and aligned with the natural center of attention. 
\textit{Horizontal} performed the worst overall, likely because the top-down projection removes the vertical dimension and forces spatial interpretation to rely more heavily on depth perception, making it harder for participants to estimate and communicate spatial relationships effectively.

\smallskip
\noindent\textbf{Layout curvature increased spatial referencing difficulty.}
Overall, we found strong evidence supporting our hypothesis~\textbf{\ref{h2}}.
Across tasks and projections, increasing layout curvature was associated with poorer performance, with \textbf{Flat} yielding the fastest completion times and lowest error counts overall, and \textbf{Full-Sphere} producing the highest error counts.
The distinction between Hemisphere and Full-Sphere was more pronounced for error count than for completion time.

These findings are consistent with prior work by Liu et al.~\cite{liu_effects_2022, liu_design_2020}, who found that flat layouts supported better spatial memory performance than more curved wraparound layouts in single-user VR. 
Our results extend this finding to collaborative cross-reality settings by showing that increasing layout curvature also makes spatial referencing more difficult when users must coordinate across VR and PC views.
In particular, as curvature increases, object layouts extend farther around the VR user, making it more difficult for collaborators to establish and maintain a shared spatial frame across the VR and PC views.
This reduces \textit{spatial congruency} between collaborators by making simple front-facing heuristics less reliable and increasing ambiguity in target localization and reference alignment.
 
\smallskip
\noindent\textbf{Equirectangular projection remained robust as layout curvature increased.} 
We can overall confirm our hypothesis~\textbf{\ref{h3}} based on analyses of completion time, error count, and subjective ratings, as the advantage of \textit{equirectangular} over the planar projections became more pronounced with increasing layout curvature, with the largest gap observed in the \textbf{Full-Sphere} layouts.
These results suggest that \textit{equirectangular} remained robust as task difficulty increased with curvature, whereas the planar projections became less reliable under these layout geometries.

In \textbf{Flat} layouts, projection differences were present but relatively limited, and \textit{vertical} often remained competitive with \textit{equirectangular}. 
This pattern is expected given the nature of the planar projections. 
For \textit{vertical}, the projection plane is aligned with the same frontal structure as the layout itself, producing an almost direct copy of the 3D layout in the 2D view. 
By contrast, \textit{equirectangular} introduces geometric distortions when mapping the layout into a panoramic representation. 
Despite this, we did not observe any significant performance degradation associated with this distortion, suggesting that it was not substantial enough to interfere with the spatial cues most relevant to the task.

As layouts became more curved, however, the performance gap between projections became more pronounced, and each planar projection exhibited degradation under specific layout conditions.
Each planar method was most affected when the layout geometry directly exposed the representational limitations of its mapping.
\textit{Horizontal} degraded most strongly in the \textbf{Hemisphere} layout because its top-down mapping collapses the vertical dimension, thereby discarding height information that remains important for spatial referencing in this layout and that the other projections were able to preserve.
\textit{PCA} degraded most strongly in the \textbf{Full-Sphere} layout because its projection plane is determined by the dominant variance structure of the point distribution. 
When objects surround the user more uniformly, the first two principal components are no longer anchored to a clear frontal organization, so the resulting plane can become diagonal and visually arbitrary as it optimizes the 3D-to-2D mapping to reduce overlap. 
Although this may be geometrically optimized, it often yields a view that is confusing and difficult for participants to interpret in spatial referencing tasks.
\textit{Vertical} remained relatively comparable to \textit{equirectangular}, possibly because its frontal organization better matches participants' screen-based spatial expectations.
This makes the representation relatively intuitive in conditions where the object layout remains primarily frontal.
However, it became less effective as objects extended farther around the user.
In these cases, points near the far edges of the surrounding field, including those behind the user, could project onto positions very close to points directly in front, which sometimes introduced ambiguity in communication between partners in our observations.
Borowski et al.\ likewise identified overlap as a limitation of planar projections in a cross-reality collaboration setting~\cite{borowski_spatialstrates_2025}.

The interaction between \textit{equirectangular} projection and layout curvature suggests that whether the spatial properties of a VR layout can be preserved and conveyed to PC collaborators in a spatially congruent manner depends fundamentally on how well the projection maintains the surrounding spatial structure. \textit{Equirectangular} preserved the full angular organization of the layout within a panoramic representation, allowing collaborators on both platforms to interpret object locations within a shared and continuously synchronized reference frame. 
This likely supported more spatially congruent cross-view alignment by preserving left--right and around-the-user relationships even as the layout became more curved. 
More broadly, these findings indicate that \textit{equirectangular}, and potentially spherical projections more generally, may be better suited for collaborative cross-reality tasks, particularly those involving surrounding layouts in an immersive environment and high spatial referencing demands.

\smallskip
\noindent\textbf{As task complexity increases, coordination draws on factors beyond projection.}
We did not find conclusive support for \textbf{\ref{h4}} in either the quantitative results or our observations.
Projection choice appeared most consequential during initial grounding, when collaborators had to establish a shared frame of reference directly from the mapping between the VR and PC views. 
In Task~1, where one-way referential communication depended most heavily on this mapping, \textit{equirectangular} showed the clearest objective advantage, particularly under curved layouts. 
However, this advantage did not increase with task complexity as hypothesized: differences became more selective in Task~2 and were often no longer significant in Task~3. 
This pattern suggests that projection primarily shaped initial grounding, whereas sustained coordination increasingly drew on shared anchors and common ground developed through interaction.
More generally, spatial referencing in complex workflows likely depends on factors beyond projection geometry, which we discuss further in~\autoref{limitations}.
\section{Limitations and Future Work}
\label{limitations}
Several limitations should be acknowledged when interpreting these findings, and they also inform directions for future research.

\smallskip
\noindent\textbf{Other Factors in Spatial Referencing.}
Spatial referencing depends on factors beyond projection, including semantic content, object density, participant roles, and awareness cues.
Within the ``eyes-and-shoes'' framework~\cite{saffo_through_2023}, our study focused specifically on perspective sharing: how collaborators understand the same space across asymmetric VR and PC views.
We therefore used abstract targets, static layouts, fixed device roles, co-located dyads, and minimal awareness cues beyond synchronized task feedback and a shared landmark.
These deliberate choices reduced semantic, dynamic, device-adaptation, network, and auxiliary coordination confounds, allowing us to isolate projection and layout geometry.
However, they also limit the ecological validity of our findings.
Future work should examine projection techniques in more realistic workflows involving semantically meaningful content, varying object densities, dynamic environments and viewpoints, role switching, remote collaboration, and richer coordination support such as gaze cues, shared pointers, view indicators, and annotations.

\smallskip
\noindent\textbf{Layout and Projection Scope.}
Our study examined curvature-based layouts, while other spatial organizations observed in immersive analytics, such as clusters and layered depth arrangements, may produce different projection trade-offs.
We also evaluated only one spherical projection, \textit{equirectangular}, because it provides a simple, screen-friendly mapping to a rectangular 2D view.
Other spherical projections may offer different continuity--distortion trade-offs~\cite{chen_gansda_2022} and should be examined in future cross-reality collaboration studies.
\section{Conclusion}
In this paper, we presented \textsc{CrossAtlas}, a cross-reality collaboration platform for evaluating how different bidirectional projection techniques shape spatial referencing between VR and PC collaborators.
Through a controlled study with 24 dyads, we showed that projection choice is a critical design factor in collaborative cross-reality settings: \textit{equirectangular} outperformed planar alternatives in accuracy, efficiency, and subjective experience across most scenarios, and remained especially robust as layout curvature increased. 
We further found that curved layouts increased the difficulty of spatial referencing, and that the costs of curvature depended strongly on how well a projection preserved the surrounding spatial structure. 
Taken together, these findings suggest that \textit{equirectangular}, and potentially spherical projections more broadly, could be a promising design direction for collaborative cross-reality tasks involving surrounding layouts and high spatial referencing demands.
 
\acknowledgments{This research was supported in part by NSF award IIS-2441310.}

\bibliographystyle{abbrv-doi-hyperref}

\bibliography{references}

@misc{yjs,
  author       = "{Kevin Jahns and Yjs contributors}",
  title        = "{Yjs}",
  year         = {2026},
  howpublished = {\url{https://yjs.dev/}},
}

@misc{anujs,
  author       = "{JPMorganChase and Anu contributors}",
  title        = "{Anu}",
  year         = {2025},
  howpublished = {\url{https://jpmorganchase.github.io/anu/}},
}

@misc{babylonjs,
  author       = "{David Catuhe and Babylon.js contributors}",
  title        = "{Babylon.js}",
  year         = {2013},
  howpublished = {\url{https://www.babylonjs.com}},
}

@misc{webxr,
  author       = "{Immersive Web Working Group}",
  title        = "{WebXR}",
  year         = {2024},
  howpublished = {\url{https://www.w3.org/TR/webxr}},
}

@article{ens_revisiting_2019,
    title = {Revisiting collaboration through mixed reality: {The} evolution of groupware},
    volume = {131},
    issn = {1071-5819},
    shorttitle = {Revisiting collaboration through mixed reality},
    url = {https://doi.org/10.1016/j.ijhcs.2019.05.011},
    doi = {10.1016/j.ijhcs.2019.05.011},
    number = {C},
    urldate = {2026-02-15},
    journal = {Int. J. Hum.-Comput. Stud.},
    author = {Ens, Barrett and Lanir, Joel and Tang, Anthony and Bateman, Scott and Lee, Gun and Piumsomboon, Thammathip and Billinghurst, Mark},
    month = nov,
    year = {2019},
    pages = {81--98},
}

@article{milgram_taxonomy_1994,
    title = {A {Taxonomy} of {Mixed} {Reality} {Visual} {Displays}},
    volume = {E77-D},
    url = {https://globals.ieice.org/en_transactions/information/10.1587/e77-d_12_1321/#},
    language = {en},
    number = {12},
    urldate = {2026-02-15},
    journal = {IEICE Transactions on Information},
    publisher = {The Institute of Electronics, Information and Communication Engineers},
    author = {Milgram, Paul and Kishino, Fumio},
    month = dec,
    year = {1994},
    pages = {1321--1329},
}

@article{lee_shared_2021,
    title = {Shared {Surfaces} and {Spaces}: {Collaborative} {Data} {Visualisation} in a {Co}-located {Immersive} {Environment}},
    volume = {27},
    issn = {1941-0506},
    shorttitle = {Shared {Surfaces} and {Spaces}},
    url = {https://ieeexplore.ieee.org/abstract/document/9222346},
    doi = {10.1109/TVCG.2020.3030450},
    number = {2},
    urldate = {2025-05-04},
    journal = {IEEE Transactions on Visualization and Computer Graphics},
    author = {Lee, Benjamin and Hu, Xiaoyun and Cordeil, Maxime and Prouzeau, Arnaud and Jenny, Bernhard and Dwyer, Tim},
    month = feb,
    year = {2021},
    pages = {1171--1181},
}

@article{cordeil_immersive_2017,
    title = {Immersive {Collaborative} {Analysis} of {Network} {Connectivity}: {CAVE}-style or {Head}-{Mounted} {Display}?},
    volume = {23},
    issn = {1077-2626},
    shorttitle = {Immersive {Collaborative} {Analysis} of {Network} {Connectivity}},
    url = {http://ieeexplore.ieee.org/document/7539620/},
    doi = {10.1109/TVCG.2016.2599107},
    language = {en},
    number = {1},
    urldate = {2022-04-14},
    journal = {IEEE Transactions on Visualization and Computer Graphics},
    author = {Cordeil, Maxime and Dwyer, Tim and Klein, Karsten and Laha, Bireswar and Marriott, Kim and Thomas, Bruce H.},
    month = jan,
    year = {2017},
    pages = {441--450},
}

@article{bates2015fitting,
  title={Fitting linear mixed-effects models using lme4},
  author={Bates, Douglas and M{\"a}chler, Martin and Bolker, Ben and Walker, Steve},
  journal={Journal of statistical software},
  volume={67},
  pages={1--48},
  year={2015},
  doi={10.18637/jss.v067.i01}
}

@article{bolker2009generalized,
  title={Generalized linear mixed models: a practical guide for ecology and evolution},
  author={Bolker, Benjamin M and Brooks, Mollie E and Clark, Connie J and Geange, Shane W and Poulsen, John R and Stevens, M Henry H and White, Jada-Simone S},
  journal={Trends in ecology \& evolution},
  volume={24},
  number={3},
  pages={127--135},
  year={2009},
  publisher={Elsevier},
  doi={10.1016/j.tree.2008.10.008}
}

@article{lenth2016least,
  title={Least-squares means: the R package lsmeans},
  author={Lenth, Russell V},
  journal={Journal of statistical software},
  volume={69},
  pages={1--33},
  year={2016},
  doi={10.18637/jss.v069.i01}
}

@book{field2012discovering,
  title={Discovering statistics using R},
  author={Field, Andy and Field, Zoe and Miles, Jeremy},
  year={2012},
  publisher={SAGE Publications},
  isbn={9781446200469}
}

@book{hilbe2011negative,
  title={Negative binomial regression},
  author={Hilbe, Joseph M},
  year={2011},
  publisher={Cambridge University Press},
  isbn={9781139011488}
}

@inproceedings{lisle_evaluating_2020,
    title = {Evaluating the {Benefits} of the {Immersive} {Space} to {Think}},
    url = {https://ieeexplore.ieee.org/abstract/document/9090620},
    doi = {10.1109/VRW50115.2020.00073},
    urldate = {2024-10-16},
    booktitle = {2020 {IEEE} {Conference} on {Virtual} {Reality} and {3D} {User} {Interfaces} {Abstracts} and {Workshops} ({VRW})},
    author = {Lisle, Lee and Chen, Xiaoyu and Edward Gitre, J.K. and North, Chris and Bowman, Doug A.},
    month = mar,
    year = {2020},
    pages = {331--337},
}

@article{yang_litforager_2025,
    title = {{LitForager}: {Exploring} {Multimodal} {Literature} {Foraging} {Strategies} in {Immersive} {Sensemaking}},
    volume = {31},
    copyright = {https://ieeexplore.ieee.org/Xplorehelp/downloads/license-information/IEEE.html},
    issn = {1077-2626, 1941-0506, 2160-9306},
    shorttitle = {Litforager},
    url = {https://ieeexplore.ieee.org/document/11192754/},
    doi = {10.1109/TVCG.2025.3616732},
    number = {11},
    urldate = {2026-01-20},
    journal = {IEEE Transactions on Visualization and Computer Graphics},
    author = {Yang, Haoyang and Faa, Elliott H. and Liu, Weijian and Guo, Shunan and Chau, Duen Horng and Yang, Yalong},
    month = nov,
    year = {2025},
    pages = {9614--9624},
}

@inproceedings{liu_investigating_2024,
    title = {Investigating the {Effects} of {Physical} {Landmarks} on {Spatial} {Memory} for {Information} {Visualisation} in {Augmented} {Reality}},
    issn = {2473-0726},
    url = {https://ieeexplore.ieee.org/document/10765423},
    doi = {10.1109/ISMAR62088.2024.00043},
    urldate = {2025-08-18},
    booktitle = {2024 {IEEE} {International} {Symposium} on {Mixed} and {Augmented} {Reality} ({ISMAR})},
    author = {Liu, Jiazhou and Satriadi, Kadek Ananta and Ens, Barrett and Dwyer, Tim},
    month = oct,
    year = {2024},
    pages = {289--298},
}

@inproceedings{horak_when_2018,
    address = {Montreal QC, Canada},
    title = {When {David} {Meets} {Goliath}: {Combining} {Smartwatches} with a {Large} {Vertical} {Display} for {Visual} {Data} {Exploration}},
    isbn = {978-1-4503-5620-6},
    shorttitle = {When {David} {Meets} {Goliath}},
    url = {http://dl.acm.org/citation.cfm?doid=3173574.3173593},
    doi = {10.1145/3173574.3173593},
    language = {en},
    urldate = {2020-08-12},
    booktitle = {Proceedings of the 2018 {CHI} {Conference} on {Human} {Factors} in {Computing} {Systems}  - {CHI} '18},
    publisher = {ACM Press},
    author = {Horak, Tom and Badam, Sriram Karthik and Elmqvist, Niklas and Dachselt, Raimund},
    year = {2018},
    pages = {1--13},
}

@inproceedings{tong_towards_2023,
    title = {Towards an {Understanding} of {Distributed} {Asymmetric} {Collaborative} {Visualization} on {Problem}-solving},
    issn = {2642-5254},
    url = {https://ieeexplore.ieee.org/document/10108427},
    doi = {10.1109/VR55154.2023.00054},
    urldate = {2025-05-04},
    booktitle = {2023 {IEEE} {Conference} {Virtual} {Reality} and {3D} {User} {Interfaces} ({VR})},
    author = {Tong, Wai and Xia, Meng and Wong, Kam Kwai and Bowman, Doug A. and Pong, Ting-Chuen and Qu, Huamin and Yang, Yalong},
    month = mar,
    year = {2023},
    pages = {387--397},
}

@inproceedings{saffo_through_2023,
    address = {New York, NY, USA},
    series = {{CHI} '23},
    title = {Through {Their} {Eyes} and {In} {Their} {Shoes}: {Providing} {Group} {Awareness} {During} {Collaboration} {Across} {Virtual} {Reality} and {Desktop} {Platforms}},
    isbn = {978-1-4503-9421-5},
    shorttitle = {Through {Their} {Eyes} and {In} {Their} {Shoes}},
    url = {https://dl.acm.org/doi/10.1145/3544548.3581093},
    doi = {10.1145/3544548.3581093},
    urldate = {2025-05-04},
    booktitle = {Proceedings of the 2023 {CHI} {Conference} on {Human} {Factors} in {Computing} {Systems}},
    publisher = {Association for Computing Machinery},
    author = {Saffo, David and Batch, Andrea and Dunne, Cody and Elmqvist, Niklas},
    month = apr,
    year = {2023},
    pages = {1--15},
}

@misc{srinivasan_heedvision_2025,
    title = {{HeedVision}: {Attention} {Awareness} in {Collaborative} {Immersive} {Analytics} {Environments}},
    shorttitle = {{HeedVision}},
    url = {http://arxiv.org/abs/2505.07069},
    doi = {10.48550/arXiv.2505.07069},
    urldate = {2025-05-26},
    publisher = {arXiv},
    author = {Srinivasan, Arvind and Elmqvist, Niklas},
    month = may,
    year = {2025},
    note = {arXiv:2505.07069 [cs]},
}

@article{borowski_dashspace_2025,
    title = {{DashSpace}: {A} {Live} {Collaborative} {Platform} for {Immersive} and {Ubiquitous} {Analytics}},
    issn = {1941-0506},
    shorttitle = {{DashSpace}},
    url = {https://ieeexplore.ieee.org/abstract/document/10869395},
    doi = {10.1109/TVCG.2025.3537679},
    urldate = {2025-05-04},
    journal = {IEEE Transactions on Visualization and Computer Graphics},
    author = {Borowski, Marcel and Butcher, Peter W. S. and Kristensen, Janus Bager and Petersen, Jonas Oxenbøll and Ritsos, Panagiotis D. and Klokmose, Clemens N. and Elmqvist, Niklas},
    year = {2025},
    pages = {1--13},
}

@article{seraji_analyzing_2024,
    title = {Analyzing {User} {Behaviour} {Patterns} in a {Cross}-{Virtuality} {Immersive} {Analytics} {System}},
    volume = {30},
    issn = {1941-0506},
    url = {https://ieeexplore.ieee.org/abstract/document/10471345},
    doi = {10.1109/TVCG.2024.3372129},
    number = {5},
    urldate = {2025-05-04},
    journal = {IEEE Transactions on Visualization and Computer Graphics},
    author = {Seraji, Mohammad Rajabi and Piray, Parastoo and Zahednejad, Vahid and Stuerzlinger, Wolfgang},
    month = may,
    year = {2024},
    pages = {2613--2623},
}

@inproceedings{borowski_spatialstrates_2025,
    address = {New York, NY, USA},
    series = {{UIST} '25},
    title = {Spatialstrates: {Cross}-{Reality} {Collaboration} through {Spatial} {Hypermedia}},
    isbn = {979-8-4007-2037-6},
    shorttitle = {Spatialstrates},
    url = {https://dl.acm.org/doi/10.1145/3746059.3747708},
    doi = {10.1145/3746059.3747708},
    urldate = {2025-10-01},
    booktitle = {Proceedings of the 38th {Annual} {ACM} {Symposium} on {User} {Interface} {Software} and {Technology}},
    publisher = {Association for Computing Machinery},
    author = {Borowski, Marcel and Grønbæk, Jens Emil Sloth and Butcher, Peter W. S. and Ritsos, Panagiotis D. and Klokmose, Clemens Nylandsted and Elmqvist, Niklas},
    month = sep,
    year = {2025},
    pages = {1--14},
}

@inproceedings{gutwin_group_2004,
    address = {New York, NY, USA},
    series = {{CSCW} '04},
    title = {Group awareness in distributed software development},
    isbn = {978-1-58113-810-8},
    url = {https://dl.acm.org/doi/10.1145/1031607.1031621},
    doi = {10.1145/1031607.1031621},
    urldate = {2026-03-15},
    booktitle = {Proceedings of the 2004 {ACM} conference on {Computer} supported cooperative work},
    publisher = {Association for Computing Machinery},
    author = {Gutwin, Carl and Penner, Reagan and Schneider, Kevin},
    month = nov,
    year = {2004},
    pages = {72--81},
}

@inproceedings{badam_polychrome_2014,
    address = {New York, NY, USA},
    series = {{ITS} '14},
    title = {{PolyChrome}: {A} {Cross}-{Device} {Framework} for {Collaborative} {Web} {Visualization}},
    isbn = {978-1-4503-2587-5},
    shorttitle = {{PolyChrome}},
    url = {https://dl.acm.org/doi/10.1145/2669485.2669518},
    doi = {10.1145/2669485.2669518},
    urldate = {2026-03-15},
    booktitle = {Proceedings of the {Ninth} {ACM} {International} {Conference} on {Interactive} {Tabletops} and {Surfaces}},
    publisher = {Association for Computing Machinery},
    author = {Badam, Sriram Karthik and Elmqvist, Niklas},
    month = nov,
    year = {2014},
    pages = {109--118},
}

@inproceedings{kim_hugin_2010,
    address = {New York, NY, USA},
    series = {{ITS} '10},
    title = {Hugin: a framework for awareness and coordination in mixed-presence collaborative information visualization},
    isbn = {978-1-4503-0399-6},
    shorttitle = {Hugin},
    url = {https://dl.acm.org/doi/10.1145/1936652.1936694},
    doi = {10.1145/1936652.1936694},
    urldate = {2026-03-15},
    booktitle = {{ACM} {International} {Conference} on {Interactive} {Tabletops} and {Surfaces}},
    publisher = {Association for Computing Machinery},
    author = {Kim, KyungTae and Javed, Waqas and Williams, Cary and Elmqvist, Niklas and Irani, Pourang},
    month = nov,
    year = {2010},
    pages = {231--240},
}

@article{billinghurst_collaborative_2002,
    title = {Collaborative augmented reality},
    volume = {45},
    issn = {00010782},
    url = {http://portal.acm.org/citation.cfm?doid=514236.514265},
    doi = {10.1145/514236.514265},
    number = {7},
    urldate = {2020-07-09},
    journal = {Communications of the ACM},
    author = {Billinghurst, Mark and Kato, Hirokazu},
    month = jul,
    year = {2002},
}

@inproceedings{butscher_clusters_2018,
    address = {New York, NY, USA},
    series = {{CHI} '18},
    title = {Clusters, {Trends}, and {Outliers}: {How} {Immersive} {Technologies} {Can} {Facilitate} the {Collaborative} {Analysis} of {Multidimensional} {Data}},
    isbn = {978-1-4503-5620-6},
    shorttitle = {Clusters, {Trends}, and {Outliers}},
    url = {https://dl.acm.org/doi/10.1145/3173574.3173664},
    doi = {10.1145/3173574.3173664},
    urldate = {2025-05-04},
    booktitle = {Proceedings of the 2018 {CHI} {Conference} on {Human} {Factors} in {Computing} {Systems}},
    publisher = {Association for Computing Machinery},
    author = {Butscher, Simon and Hubenschmid, Sebastian and Müller, Jens and Fuchs, Johannes and Reiterer, Harald},
    month = apr,
    year = {2018},
    pages = {1--12},
}

@inproceedings{schroder_collaborating_2023,
    address = {New York, NY, USA},
    series = {{CHI} '23},
    title = {Collaborating {Across} {Realities}: {Analytical} {Lenses} for {Understanding} {Dyadic} {Collaboration} in {Transitional} {Interfaces}},
    isbn = {978-1-4503-9421-5},
    shorttitle = {Collaborating {Across} {Realities}},
    url = {https://dl.acm.org/doi/10.1145/3544548.3580879},
    doi = {10.1145/3544548.3580879},
    urldate = {2025-05-04},
    booktitle = {Proceedings of the 2023 {CHI} {Conference} on {Human} {Factors} in {Computing} {Systems}},
    publisher = {Association for Computing Machinery},
    author = {Schröder, Jan-Henrik and Schacht, Daniel and Peper, Niklas and Hamurculu, Anita Marie and Jetter, Hans-Christian},
    month = apr,
    year = {2023},
    pages = {1--16},
}

@article{liu_effects_2022,
    title = {Effects of {Display} {Layout} on {Spatial} {Memory} for {Immersive} {Environments}},
    volume = {6},
    url = {https://dl.acm.org/doi/10.1145/3567729},
    doi = {10.1145/3567729},
    number = {ISS},
    urldate = {2025-05-04},
    journal = {Proc. ACM Hum.-Comput. Interact.},
    author = {Liu, Jiazhou and Prouzeau, Arnaud and Ens, Barrett and Dwyer, Tim},
    month = nov,
    year = {2022},
    pages = {576:468--576:488},
}

@inproceedings{liu_design_2020,
    title = {Design and {Evaluation} of {Interactive} {Small} {Multiples} {Data} {Visualisation} in {Immersive} {Spaces}},
    issn = {2642-5254},
    url = {https://ieeexplore.ieee.org/abstract/document/9089546},
    doi = {10.1109/VR46266.2020.00081},
    urldate = {2025-05-04},
    booktitle = {2020 {IEEE} {Conference} on {Virtual} {Reality} and {3D} {User} {Interfaces} ({VR})},
    author = {Liu, Jiazhou and Prouzeau, Arnaud and Ens, Barrett and Dwyer, Tim},
    month = mar,
    year = {2020},
    pages = {588--597},
}

@article{heer_design_2008,
    title = {Design considerations for collaborative visual analytics},
    volume = {7},
    issn = {1473-8716},
    url = {https://doi.org/10.1145/1391107.1391112},
    doi = {10.1145/1391107.1391112},
    number = {1},
    urldate = {2026-03-16},
    journal = {Information Visualization},
    author = {Heer, Jeffrey and Agrawala, Maneesh},
    month = mar,
    year = {2008},
    pages = {49--62},
}

@incollection{clark_grounding_1991,
    address = {Washington, DC, US},
    title = {Grounding in communication},
    isbn = {978-1-55798-121-9},
    url = {https://content.apa.org/books/10096-006},
    doi = {10.1037/10096-006},
    language = {en},
    urldate = {2026-03-16},
    booktitle = {Perspectives on socially shared cognition},
    publisher = {American Psychological Association},
    author = {Clark, Herbert H. and Brennan, Susan E.},
    editor = {Resnick, Lauren B. and Levine, John M. and Teasley, Stephanie D.},
    year = {1991},
    pages = {127--149},
}

@inproceedings{wong_spatial_2025,
    address = {New York, NY, USA},
    series = {{CHI} '25},
    title = {Spatial {Heterogeneity} in {Distributed} {Mixed} {Reality} {Collaboration}},
    isbn = {979-8-4007-1394-1},
    url = {https://dl.acm.org/doi/10.1145/3706598.3714033},
    doi = {10.1145/3706598.3714033},
    urldate = {2026-03-15},
    booktitle = {Proceedings of the 2025 {CHI} {Conference} on {Human} {Factors} in {Computing} {Systems}},
    publisher = {Association for Computing Machinery},
    author = {Wong, Emily and Genay, Adélaïde and Grønbæk, Jens Emil Sloth and Velloso, Eduardo},
    month = apr,
    year = {2025},
    pages = {1--19},
}

@inproceedings{wang_mrtransformer_2024,
    title = {{MRTransformer}: {Transforming} {Avatar} {Non}-verbal {Behavior} for {Remote} {MR} {Collaboration} in {Incongruent} {Spaces}},
    issn = {2771-1110},
    shorttitle = {{MRTransformer}},
    url = {https://ieeexplore.ieee.org/document/10765238},
    doi = {10.1109/ISMAR-Adjunct64951.2024.00157},
    urldate = {2026-03-16},
    booktitle = {2024 {IEEE} {International} {Symposium} on {Mixed} and {Augmented} {Reality} {Adjunct} ({ISMAR}-{Adjunct})},
    author = {Wang, Cheng Yao and Kim, Hyunju and Panda, Payod and Ofek, Eyal and Franco, Mar Gonzalez and Won, Andrea Stevenson},
    month = oct,
    year = {2024},
    note = {ISSN: 2771-1110},
    pages = {545--548},
}

@article{huang_surfshare_2024,
    title = {{SurfShare}: {Lightweight} {Spatially} {Consistent} {Physical} {Surface} and {Virtual} {Replica} {Sharing} with {Head}-mounted {Mixed}-{Reality}},
    volume = {7},
    shorttitle = {{SurfShare}},
    url = {https://dl.acm.org/doi/10.1145/3631418},
    doi = {10.1145/3631418},
    number = {4},
    urldate = {2026-03-16},
    journal = {Proc. ACM Interact. Mob. Wearable Ubiquitous Technol.},
    author = {Huang, Xincheng and Xiao, Robert},
    month = jan,
    year = {2024},
    pages = {162:1--162:24},
}

@article{yang_origin-destination_2019,
    title = {Origin-{Destination} {Flow} {Maps} in {Immersive} {Environments}},
    volume = {25},
    issn = {1941-0506},
    url = {https://ieeexplore.ieee.org/document/8440844},
    doi = {10.1109/TVCG.2018.2865192},
    number = {1},
    urldate = {2026-03-16},
    journal = {IEEE Transactions on Visualization and Computer Graphics},
    author = {Yang, Yalong and Dwyer, Tim and Jenny, Bernhard and Marriott, Kim and Cordeil, Maxime and Chen, Haohui},
    month = jan,
    year = {2019},
    pages = {693--703},
}

@article{zhao_spatialtouch_2025,
    title = {{SpatialTouch}: {Exploring} {Spatial} {Data} {Visualizations} in {Cross}-{Reality}},
    volume = {31},
    issn = {1941-0506},
    shorttitle = {{SpatialTouch}},
    url = {https://ieeexplore.ieee.org/abstract/document/10670539},
    doi = {10.1109/TVCG.2024.3456368},
    number = {1},
    urldate = {2025-05-04},
    journal = {IEEE Transactions on Visualization and Computer Graphics},
    author = {Zhao, Lixiang and Isenberg, Tobias and Xie, Fuqi and Liang, Hai-Ning and Yu, Lingyun},
    month = jan,
    year = {2025},
    pages = {897--907},
}

@misc{lu_design_2025,
    title = {A {Design} {Space} for {Visualization} {Transitions} of {3D} {Spatial} {Data} in {Hybrid} {AR}-{Desktop} {Environments}},
    url = {http://arxiv.org/abs/2506.22250},
    doi = {10.48550/arXiv.2506.22250},
    urldate = {2025-07-22},
    publisher = {arXiv},
    author = {Lu, Yucheng and Rau, Tobias and Lee, Benjamin and Köhn, Andreas and Sedlmair, Michael and Sandor, Christian and Isenberg, Tobias},
    month = jun,
    year = {2025},
    note = {arXiv:2506.22250 [cs]},
}

@techreport{snyder_map_1987,
    title = {Map projections: {A} working manual},
    issn = {2330-7102},
    shorttitle = {Map projections},
    url = {https://pubs.usgs.gov/publication/pp1395},
    doi = {10.3133/pp1395},
    language = {en},
    number = {1395},
    urldate = {2026-03-16},
    institution = {U.S. Government Printing Office},
    author = {Snyder, John P.},
    year = {1987},
    note = {Publication Title: Professional Paper},
}

@inproceedings{lee_design_2022,
    address = {New York, NY, USA},
    series = {{CHI} '22},
    title = {A {Design} {Space} {For} {Data} {Visualisation} {Transformations} {Between} {2D} {And} {3D} {In} {Mixed}-{Reality} {Environments}},
    isbn = {978-1-4503-9157-3},
    url = {https://dl.acm.org/doi/10.1145/3491102.3501859},
    doi = {10.1145/3491102.3501859},
    urldate = {2024-09-23},
    booktitle = {Proceedings of the 2022 {CHI} {Conference} on {Human} {Factors} in {Computing} {Systems}},
    publisher = {Association for Computing Machinery},
    author = {Lee, Benjamin and Cordeil, Maxime and Prouzeau, Arnaud and Jenny, Bernhard and Dwyer, Tim},
    month = apr,
    year = {2022},
    pages = {1--14},
}

@article{ma_international_1998,
    title = {The {International} {Celestial} {Reference} {Frame} as {Realized} by {VeryLong} {Baseline} {Interferometry}},
    volume = {116},
    issn = {1538-3881},
    url = {https://iopscience.iop.org/article/10.1086/300408},
    doi = {10.1086/300408},
    language = {en},
    number = {1},
    urldate = {2026-03-16},
    journal = {The Astronomical Journal},
    publisher = {IOP Publishing},
    author = {Ma, C. and Arias, E. F. and Eubanks, T. M. and Fey, A. L. and Gontier, A.-M. and Jacobs, C. S. and Sovers, O. J. and Archinal, B. A. and Charlot, P.},
    month = jul,
    year = {1998},
    pages = {516},
}

@article{hussain_evaluation_2021,
    title = {Evaluation of 360° {Image} {Projection} {Formats}; {Comparing} {Format} {Conversion} {Distortion} {Using} {Objective} {Quality} {Metrics}},
    volume = {7},
    issn = {2313-433X},
    url = {https://pmc.ncbi.nlm.nih.gov/articles/PMC8404912/},
    doi = {10.3390/jimaging7080137},
    number = {8},
    urldate = {2026-03-16},
    journal = {Journal of Imaging},
    author = {Hussain, Ikram and Kwon, Oh-Jin},
    month = aug,
    year = {2021},
    pages = {137},
}

@article{shafi_360-degree_2020,
    title = {360-{Degree} {Video} {Streaming}: {A} {Survey} of the {State} of the {Art}},
    volume = {12},
    shorttitle = {360-{Degree} {Video} {Streaming}},
    doi = {10.3390/sym12091491},
    journal = {Symmetry},
    author = {Shafi, Rabia and Shuai, Wan and Younus, Muhammad},
    month = sep,
    year = {2020},
    pages = {1491},
}

@inproceedings{corbillon_viewport-adaptive_2017,
    title = {Viewport-{Adaptive} {Navigable} 360-{Degree} {Video} {Delivery}},
    url = {http://arxiv.org/abs/1609.08042},
    doi = {10.1109/ICC.2017.7996611},
    urldate = {2026-03-16},
    booktitle = {2017 {IEEE} {International} {Conference} on {Communications} ({ICC})},
    author = {Corbillon, Xavier and Simon, Gwendal and Devlic, Alisa and Chakareski, Jacob},
    month = may,
    year = {2017},
    note = {arXiv:1609.08042 [cs]},
    pages = {1--7},
}

@article{batch_there_2020,
    title = {There {Is} {No} {Spoon}: {Evaluating} {Performance}, {Space} {Use}, and {Presence} with {Expert} {Domain} {Users} in {Immersive} {Analytics}},
    volume = {26},
    issn = {1941-0506},
    shorttitle = {There {Is} {No} {Spoon}},
    url = {https://ieeexplore.ieee.org/document/8820171},
    doi = {10.1109/TVCG.2019.2934803},
    number = {1},
    urldate = {2026-03-16},
    journal = {IEEE Transactions on Visualization and Computer Graphics},
    author = {Batch, Andrea and Cunningham, Andrew and Cordeil, Maxime and Elmqvist, Niklas and Dwyer, Tim and Thomas, Bruce H. and Marriott, Kim},
    month = jan,
    year = {2020},
    pages = {536--546},
}

@inproceedings{lisle_sensemaking_2021,
    title = {Sensemaking {Strategies} with {Immersive} {Space} to {Think}},
    issn = {2642-5254},
    url = {https://ieeexplore.ieee.org/document/9417736},
    doi = {10.1109/VR50410.2021.00077},
    urldate = {2025-01-17},
    booktitle = {2021 {IEEE} {Virtual} {Reality} and {3D} {User} {Interfaces} ({VR})},
    author = {Lisle, Lee and Davidson, Kylie and Gitre, Edward J.K. and North, Chris and Bowman, Doug A.},
    month = mar,
    year = {2021},
    pages = {529--537},
}

@inproceedings{luo_where_2022,
	address = {New York, NY, USA},
	series = {{CHI} '22},
	title = {Where {Should} {We} {Put} {It}? {Layout} and {Placement} {Strategies} of {Documents} in {Augmented} {Reality} for {Collaborative} {Sensemaking}},
	isbn = {978-1-4503-9157-3},
	shorttitle = {Where {Should} {We} {Put} {It}?},
	url = {https://dl.acm.org/doi/10.1145/3491102.3501946},
	doi = {10.1145/3491102.3501946},
	urldate = {2026-03-16},
	booktitle = {Proceedings of the 2022 {CHI} {Conference} on {Human} {Factors} in {Computing} {Systems}},
	publisher = {Association for Computing Machinery},
	author = {Luo, Weizhou and Lehmann, Anke and Widengren, Hjalmar and Dachselt, Raimund},
	month = apr,
	year = {2022},
	pages = {1--16},
}

@inproceedings{johnson_you_2021,
    address = {New York, NY, USA},
    series = {{CHI} '21},
    title = {Do {You} {Really} {Need} to {Know} {Where} “{That}” {Is}? {Enhancing} {Support} for {Referencing} in {Collaborative} {Mixed} {Reality} {Environments}},
    isbn = {978-1-4503-8096-6},
    shorttitle = {Do {You} {Really} {Need} to {Know} {Where} “{That}” {Is}?},
    url = {https://dl.acm.org/doi/10.1145/3411764.3445246},
    doi = {10.1145/3411764.3445246},
    urldate = {2026-03-16},
    booktitle = {Proceedings of the 2021 {CHI} {Conference} on {Human} {Factors} in {Computing} {Systems}},
    publisher = {Association for Computing Machinery},
    author = {Johnson, Janet G and Gasques, Danilo and Sharkey, Tommy and Schmitz, Evan and Weibel, Nadir},
    month = may,
    year = {2021},
    pages = {1--14},
}

@inproceedings{muller_remote_2017,
    address = {New York, NY, USA},
    series = {{CHI} '17},
    title = {Remote {Collaboration} {With} {Mixed} {Reality} {Displays}: {How} {Shared} {Virtual} {Landmarks} {Facilitate} {Spatial} {Referencing}},
    isbn = {978-1-4503-4655-9},
    shorttitle = {Remote {Collaboration} {With} {Mixed} {Reality} {Displays}},
    url = {https://dl.acm.org/doi/10.1145/3025453.3025717},
    doi = {10.1145/3025453.3025717},
    urldate = {2026-03-16},
    booktitle = {Proceedings of the 2017 {CHI} {Conference} on {Human} {Factors} in {Computing} {Systems}},
    publisher = {Association for Computing Machinery},
    author = {Müller, Jens and Rädle, Roman and Reiterer, Harald},
    month = may,
    year = {2017},
    pages = {6481--6486},
}

@inproceedings{numan_exploring_2022,
    address = {New York, NY, USA},
    series = {{VRST} '22},
    title = {Exploring {User} {Behaviour} in {Asymmetric} {Collaborative} {Mixed} {Reality}},
    isbn = {978-1-4503-9889-3},
    url = {https://dl.acm.org/doi/10.1145/3562939.3565630},
    doi = {10.1145/3562939.3565630},
    urldate = {2026-03-16},
    booktitle = {Proceedings of the 28th {ACM} {Symposium} on {Virtual} {Reality} {Software} and {Technology}},
    publisher = {Association for Computing Machinery},
    author = {Numan, Nels and Steed, Anthony},
    month = nov,
    year = {2022},
    pages = {1--11},
}

@inproceedings{dourish_awareness_1992,
    address = {New York, NY, USA},
    series = {{CSCW} '92},
    title = {Awareness and coordination in shared workspaces},
    isbn = {978-0-89791-542-7},
    url = {https://dl.acm.org/doi/10.1145/143457.143468},
    doi = {10.1145/143457.143468},
    urldate = {2026-03-16},
    booktitle = {Proceedings of the 1992 {ACM} conference on {Computer}-supported cooperative work},
    publisher = {Association for Computing Machinery},
    author = {Dourish, Paul and Bellotti, Victoria},
    month = dec,
    year = {1992},
    pages = {107--114},
}

@article{kozhevnikov_dissociation_2001,
    title = {A dissociation between object manipulation spatial ability and spatial orientation ability},
    volume = {29},
    issn = {0090-502X},
    doi = {10.3758/bf03200477},
    language = {eng},
    number = {5},
    journal = {Memory \& Cognition},
    author = {Kozhevnikov, M. and Hegarty, M.},
    month = jul,
    year = {2001},
    pages = {745--756},
}

@article{hegarty_dissociation_2004,
    address = {Netherlands},
    title = {A dissociation between mental rotation and perspective-taking spatial abilities},
    volume = {32},
    issn = {1873-7935},
    doi = {10.1016/j.intell.2003.12.001},
    number = {2},
    journal = {Intelligence},
    publisher = {Elsevier Science},
    author = {Hegarty, Mary and Waller, David},
    year = {2004},
    pages = {175--191},
}

@article{friedman_computerized_2020,
    title = {A computerized spatial orientation test},
    volume = {52},
    issn = {1554-3528},
    doi = {10.3758/s13428-019-01277-3},
    language = {eng},
    number = {2},
    journal = {Behavior Research Methods},
    author = {Friedman, Alinda and Kohler, Bernd and Gunalp, Peri and Boone, Alexander P. and Hegarty, Mary},
    month = apr,
    year = {2020},
    pages = {799--812},
}

@article{karamazovova_spatial_2025,
    title = {Spatial perspective taking is impaired in spinocerebellar ataxias and {Friedreich} ataxia},
    volume = {15},
    issn = {2045-2322},
    url = {https://pmc.ncbi.nlm.nih.gov/articles/PMC12375724/},
    doi = {10.1038/s41598-025-16302-z},
    urldate = {2026-03-17},
    journal = {Scientific Reports},
    author = {Karamazovova, Simona and Laczó, Martina and Matuskova, Veronika and Svecova, Natalie and Stovickova, Lucie and Blichova, Zuzana and Paulasova Schwabova, Jaroslava and Kuzmiak, Michaela and Laczó, Jan and Vyhnalek, Martin},
    month = aug,
    year = {2025},
    pages = {31126},
}

@article{gunalp_spatial_2019,
    title = {Spatial perspective taking: {Effects} of social, directional, and interactive cues},
    volume = {47},
    issn = {1532-5946},
    shorttitle = {Spatial perspective taking},
    url = {https://doi.org/10.3758/s13421-019-00910-y},
    doi = {10.3758/s13421-019-00910-y},
    language = {en},
    number = {5},
    urldate = {2026-03-17},
    journal = {Memory \& Cognition},
    author = {Gunalp, Peri and Moossaian, Tara and Hegarty, Mary},
    month = jul,
    year = {2019},
    pages = {1031--1043},
}

@inproceedings{gou2026evaluating,
  title={Evaluating Replay Techniques for Asynchronous Task Handover in Immersive Analytics},
  author={Gou, Zhengtai and Long, Junxiao and Lu, Tao and Zhao, Jian and Yang, Yalong},
  booktitle={2026 IEEE Conference on Virtual Reality and 3D User Interfaces (VR)},
  pages={442--452},
  year={2026},
  organization={IEEE},
  doi={10.1109/VR67842.2026.00065}
}

@inproceedings{brehmer2026challenges,
  title={Challenges in Synchronous \& Remote Collaboration Around Visualization},
  author={Brehmer, Matthew and Cordeil, Maxime and Hurter, Christophe and Itoh, Takayuki and B{\"u}schel, Wolfgang and Jasim, Mahmood and Prouzeau, Arnaud and Saffo, David and Bartram, Lyn and Carpendale, Sheelagh and others},
  booktitle={Proceedings of the 2026 CHI Conference on Human Factors in Computing Systems},
  pages={1--17},
  doi={10.1145/3772318.3791117},
  year={2026}
}

@inproceedings{in2025exploring,
  title={Exploring Organizational Strategies in Immersive Computational Notebooks},
  author={In, Sungwon and Roy, Ayush and Krokos, Eric and Whitley, Kirsten and North, Chris and Yang, Yalong},
  booktitle={2025 IEEE International Symposium on Mixed and Augmented Reality (ISMAR)},
  pages={88--97},
  year={2025},
  organization={IEEE},
  doi={10.1109/ISMAR67309.2025.00022}
}

@article{enriquez_evaluating_2024,
  title={Evaluating layout dimensionalities in pc+vr asymmetric collaborative decision making},
  author={Enriquez, Daniel and Tong, Wai and North, Chris and Qu, Huamin and Yang, Yalong},
  journal={Proceedings of the ACM on Human-Computer Interaction},
  volume={8},
  number={ISS},
  pages={112--132},
  year={2024},
  doi={10.1145/3698130}
}

@article{yang2020embodied,
  title={Embodied navigation in immersive abstract data visualization: Is overview+ detail or zooming better for 3d scatterplots?},
  author={Yang, Yalong and Cordeil, Maxime and Beyer, Johanna and Dwyer, Tim and Marriott, Kim and Pfister, Hanspeter},
  journal={IEEE Transactions on Visualization and Computer Graphics},
  volume={27},
  number={2},
  pages={1214--1224},
  year={2020},
  publisher={IEEE},
  doi={10.1109/TVCG.2020.3030427}
}

@article{yang2018maps,
  title={Maps and globes in virtual reality},
  author={Yang, Yalong and Jenny, Bernhard and Dwyer, Tim and Marriott, Kim and Chen, Haohui and Cordeil, Maxime},
  journal={Computer Graphics Forum},
  volume={37},
  number={3},
  pages={427--438},
  year={2018},
  doi={10.1111/cgf.13431}
}

@article{hartmann_spatial_2016,
    address = {Germany},
    title = {The {Spatial} {Presence} {Experience} {Scale} ({SPES}): {A} short self-report measure for diverse media settings},
    volume = {28},
    issn = {2151-2388},
    shorttitle = {The {Spatial} {Presence} {Experience} {Scale} ({SPES})},
    doi = {10.1027/1864-1105/a000137},
    number = {1},
    journal = {Journal of Media Psychology: Theories, Methods, and Applications},
    publisher = {Hogrefe Publishing},
    author = {Hartmann, Tilo and Wirth, Werner and Schramm, Holger and Klimmt, Christoph and Vorderer, Peter and Gysbers, André and Böcking, Saskia and Ravaja, Niklas and Laarni, Jari and Saari, Timo and Gouveia, Feliz and Sacau, Ana Maria},
    year = {2016},
    pages = {1--15},
}

@article{marguet_windowspace_2025,
    title = {{WindowSpace}: {A} {Web}-{Based} {XR} {Window} {Manager} for {Interacting} with {2D} {Windows} in {Immersive} {3D} {Space}},
    volume = {9},
    shorttitle = {{WindowSpace}},
    url = {https://dl.acm.org/doi/10.1145/3773063},
    doi = {10.1145/3773063},
    number = {8},
    urldate = {2026-07-21},
    journal = {Proceedings of the ACM on Human-Computer Interaction},
    author = {Marguet, Emeric and Borowski, Marcel and Kristensen, Janus Bager and Klokmose, Clemens Nylandsted and Elmqvist, Niklas},
    month = nov,
    year = {2025},
    pages = {ISS006:119--ISS006:144},
}

@inproceedings{in_evaluating_2024,
    address = {New York, NY, USA},
    series = {{CHI} '24},
    title = {Evaluating {Navigation} and {Comparison} {Performance} of {Computational} {Notebooks} on {Desktop} and in {Virtual} {Reality}},
    isbn = {9798400703300},
    url = {https://dl.acm.org/doi/10.1145/3613904.3642932},
    doi = {10.1145/3613904.3642932},
    urldate = {2025-04-06},
    booktitle = {Proceedings of the 2024 {CHI} {Conference} on {Human} {Factors} in {Computing} {Systems}},
    publisher = {Association for Computing Machinery},
    author = {In, Sungwon and Krokos, Eric and Whitley, Kirsten and North, Chris and Yang, Yalong},
    month = may,
    year = {2024},
    pages = {1--15},
}

@inproceedings{chen_gansda_2022,
    address = {New York, NY, USA},
    series = {{CHI} '22},
    title = {{GAN}’{SDA} {Wrap}: {Geographic} {And} {Network} {Structured} {DAta} on surfaces that {Wrap} around},
    isbn = {978-1-4503-9157-3},
    shorttitle = {{GAN}’{SDA} {Wrap}},
    url = {https://dl.acm.org/doi/10.1145/3491102.3501928},
    doi = {10.1145/3491102.3501928},
    urldate = {2026-05-31},
    booktitle = {Proceedings of the 2022 {CHI} {Conference} on {Human} {Factors} in {Computing} {Systems}},
    publisher = {Association for Computing Machinery},
    author = {Chen, Kun-Ting and Dwyer, Tim and Yang, Yalong and Bach, Benjamin and Marriott, Kim},
    month = apr,
    year = {2022},
    pages = {1--16},
}
\end{document}